\documentclass[%
reprint,
superscriptaddress,
nofootinbib,
amsmath,amssymb,
aps,
prl,
]{revtex4-1}

\usepackage{graphicx} 
\usepackage{dcolumn}
\usepackage{bm} 
\usepackage{braket}
\usepackage{dsfont}
\usepackage{color,colortbl}
\usepackage[table,xcdraw]{xcolor}
\usepackage{hyperref}
\usepackage{multirow}
\usepackage{subcaption}
\usepackage{mwe}
\usepackage{booktabs}
\usepackage{caption}
\usepackage{lipsum}
\usepackage{babel,blindtext}
\usepackage[toc,page]{appendix}

\begin{document}

\title{E(3)-Equivariant Graph Neural Networks for Data-Efficient and Accurate Interatomic Potentials} 

\author{Simon Batzner$^*$}
\affiliation{John A. Paulson School of Engineering and Applied Sciences, Harvard University, Cambridge, MA 02138, USA}

\author{Albert Musaelian}
\affiliation{John A. Paulson School of Engineering and Applied Sciences, Harvard University, Cambridge, MA 02138, USA}

\author{Lixin Sun}
\affiliation{John A. Paulson School of Engineering and Applied Sciences, Harvard University, Cambridge, MA 02138, USA}

\author{Mario Geiger}
\affiliation{École Polytechnique Fédérale de Lausanne, 1015 Lausanne, Switzerland}

\author{Jonathan P. Mailoa}
\affiliation{Robert Bosch Research and Technology Center, Cambridge, MA 02139, USA}

\author{Mordechai Kornbluth}
\affiliation{Robert Bosch Research and Technology Center, Cambridge, MA 02139, USA}

\author{Nicola Molinari}
\affiliation{John A. Paulson School of Engineering and Applied Sciences, Harvard University, Cambridge, MA 02138, USA}

\author{Tess E. Smidt}
\affiliation{Computational Research Division and Center for Advanced Mathematics for Energy Research Applications, Lawrence Berkeley National Laboratory, Berkeley, CA 94720, USA}
\affiliation{Massachusetts Institute of Technology, Department of Electrical Engineering and Computer Science, Cambridge, MA 02142, USA}

\author{Boris Kozinsky$^*$}
\affiliation{John A. Paulson School of Engineering and Applied Sciences, Harvard University, Cambridge, MA 02138, USA}
\affiliation{Robert Bosch Research and Technology Center, Cambridge, MA 02139, USA}

\def\thefootnote{$^*$}\footnotetext{\textbf{Corresponding authors}\\B.K., E-mail: \url{bkoz@seas.harvard}\\ S.B., E-mail: \url{batzner@g.harvard.edu}}\def\thefootnote{\arabic{footnote}}

\begin{abstract}
This work presents Neural Equivariant Interatomic Potentials (NequIP), an E(3)-equivariant neural network approach for learning interatomic potentials from \textit{ab-initio} calculations for molecular dynamics simulations. While most contemporary symmetry-aware models use invariant convolutions and only act on scalars, NequIP employs E(3)-equivariant convolutions for interactions of geometric tensors, resulting in a more information-rich and faithful representation of atomic environments. The method achieves state-of-the-art accuracy on a challenging and diverse set of molecules and materials while exhibiting remarkable data efficiency. NequIP outperforms existing models with up to three orders of magnitude fewer training data, challenging the widely held belief that deep neural networks require massive training sets. The high data efficiency of the method allows for the construction of accurate potentials using high-order quantum chemical level of theory as reference and enables high-fidelity molecular dynamics simulations over long time scales.
\end{abstract}

\maketitle

\section{Introduction}

Molecular dynamics (MD) simulations are an indispensable tool for computational discovery in fields as diverse as energy storage, catalysis, and biological processes \cite{richards2016design, boero1998first, lindorff2011fast}. While the atomic forces required to integrate Newton's equations of motion can in principle be obtained with high fidelity from quantum-mechanical calculations such as density functional theory (DFT), in practice the unfavorable computational scaling of first-principles methods limits simulations to short time scales and small numbers of atoms. This prohibits the study of many interesting physical phenomena beyond the time and length scales that are currently accessible, even on the largest supercomputers. Owing to their simple functional form, classical models for the atomic potential energy can typically be evaluated orders of magnitude faster than first-principles methods, thereby enabling the study of large numbers of atoms over long time scales. However, due to their limited mathematical form, classical interatomic potentials, or force fields, are inherently limited in their predictive accuracy which has historically led to a fundamental trade-off between obtaining high computational efficiency while also predicting faithful dynamics of the system under study. The construction of flexible models of the interatomic potential energy based on Machine Learning (ML-IP), and in particular Neural Networks (NN-IP), has shown great promise in providing a way to move past this dilemma, promising to learn high-fidelity potentials from \textit{ab-initio} reference calculations  while retaining favorable computational efficiency \cite{behler2007generalized, gaporiginalpaper, shapeev2016moment, thompson2015spectral, vandermause2020fly, schnet_neurips, physnet_jctc, klicpera2020directional, dcf, gnnff}. One of the limiting factors of NN-IPs is that they typically require collection of large training sets of \textit{ab-initio} calculations, often including thousands or even millions of reference structures \cite{behler2007generalized, artrith2014understanding, schnet_neurips, deepmd, physnet_jctc, smith2017ani}. This computationally expensive process of training data collection has severely limited the adoption of NN-IPs as it quickly becomes a bottleneck in the development of force-fields for new systems. \\

In this work, we present the Neural Equivariant Interatomic Potential (NequIP), a highly data-efficient deep learning approach for learning interatomic potentials from reference first-principles calculations. We show that the proposed method obtains high accuracy compared to existing ML-IP methods across a wide variety of systems, including small molecules, water in different phases, an amorphous solid, a reaction at a solid/gas interface, and a Lithium superionic conductor. Furthermore, we find that NequIP exhibits exceptional data efficiency, enabling the construction of accurate interatomic potentials from limited data sets of fewer than 1,000 or even as little as 100 reference \textit{ab-initio} calculations, where other methods require orders of magnitude more. It is worth noting that on small molecular data sets, NequIP outperforms not only other neural networks, but is also competitive with kernel-based approaches, which typically obtain better predictive accuracy than NN-IPs on small data sets (although at significant additional cost scaling in training and prediction). We further demonstrate high data efficiency and accuracy with state-of-the-art results on a training set of molecular data obtained at the quantum chemical coupled-cluster level of theory. Finally, we validate the method through a series of simulations and demonstrate that we can reproduce with high fidelity structural and kinetic properties computed from NequIP simulations in comparison to \textit{ab-initio} molecular dynamics simulations (AIMD). We directly verify that the performance gains are connected with the unique E(3)-equivariant convolution architecture of the new NequIP model.\\

\subsubsection{Related Work}

\begin{figure*}[!htbp]
\includegraphics[width=\linewidth]{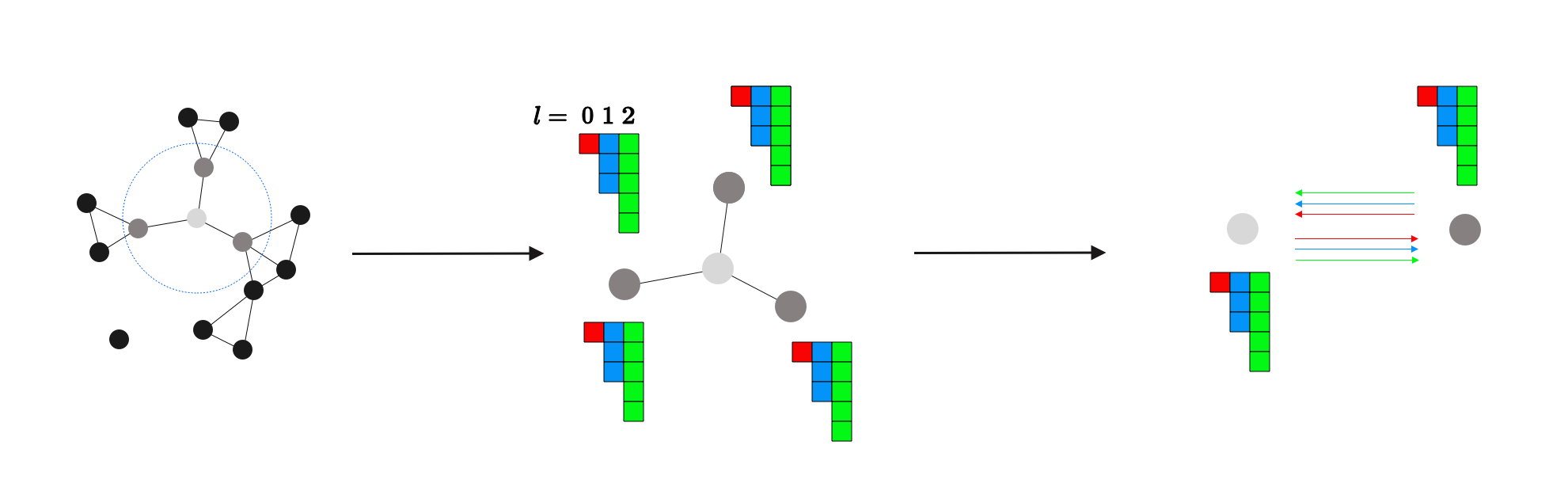}
\caption{Left: a set of atoms is interpreted as an atomic graph with local neighborhoods. Middle: every atom carries a set of scalar, vector, and higher-order tensor features with it. Right: atoms exchange information via filters, that are again geometric tensors of increasing order.}
\label{fig:neighborhoods}
\end{figure*}

The first applications of machine learning for the development of interatomic potentials were built on descriptor-based approaches combined with shallow neural networks or Gaussian Processes \cite{behler2007generalized, gaporiginalpaper}, designed to exhibit invariance with respect to translation, permutation of atoms of the same chemical species, and rotation. Recently, rotationally invariant graph neural network interatomic potentials (GNN-IPs) have emerged as a powerful architecture for deep learning of interatomic potentials that eliminates the need for hand-crafted descriptors and allows to instead learn representations on graphs of atoms from invariant features of geometric data (e.g. radial distances or angles) \cite{schnet_neurips, physnet_jctc, gnnff, klicpera2020directional}. In GNN-IPs, atomic structures are represented by collections of nodes and edges, where nodes in the graph correspond to individual atoms and edges are typically defined by simply connecting every atom to all other atoms that are closer than some cutoff distance $r_c$. Every node/atom $i$ is associated with a feature $\mathbf{h}_i \in \mathds{R}^{h}$, consisting of scalar values, which is iteratively refined via a series of convolutions over neighboring atoms $j$ based on both the distance to neighboring atoms $r_{ij}$ and their features $\mathbf{h}_j$. This iterative process allows information to be propagated along the atomic graph through a series of convolutional layers and can be viewed as a message-passing scheme \cite{gilmer2017neural}. Operating only on interatomic distances allows GNN-IPs to be rotation- and translation-invariant, making both the output as well as features internal to the network invariant to rotations. In contrast, the method outlined in this work uses relative position \textit{vectors} rather than simply distances (scalars) together with features comprised of not only scalars, but also higher-order geometric tensors. This makes internal features instead \textit{equivariant} to rotation and allows for angular information to be used by rotationally equivariant filters. Similar to other methods, we can restrict convolutions to only a local subset of all other atoms that lie closer to the central atom than a chosen cutoff distance $r_c$, see Figure \ref{fig:neighborhoods}, left.\\ 

A series of related methods have recently been proposed: DimeNet \cite{klicpera2020directional} expands on using pairwise interactions in a single convolution to include angular, three-body terms, but individual features are still comprised of scalars (distances and three-body angles are invariant to rotation), as opposed to vectors used in this work. Cormorant \cite{anderson2019cormorant} uses an equivariant neural network for property prediction on small molecules. This method is demonstrated on potential energies of small molecules but not on atomic forces or systems with periodic boundary conditions. Townshend et al. \cite{townshend2020geometric} use the framework of Tensor-Field Networks \cite{thomas2018tensor} to directly predict atomic force vectors. The predicted forces are not guaranteed by construction to conserve energy since they are not obtained as gradients of the total potential energy. This may lead to problems in simulations of molecular dynamics over long times. None of these three works \cite{klicpera2020directional, anderson2019cormorant, townshend2020geometric} demonstrates capability to perform molecular dynamics simulations.\\

After a first version of this manuscript appeared online \cite{nequip-arxiv-v1}, a series of other equivariant GNN-IPs have been proposed, such as PaiNN \cite{schutt2021equivariant} and NewtonNet \cite{haghighatlari2021newtonnet}. Both of these methods were proposed after NequIP and only make use of $l=1$ tensors. In addition, we also compare a series of other works that have since been proposed, including the GemNet \cite{klicpera2021gemnet}, SpookyNet \cite{unke2021}, and UNiTE approaches \cite{qiao2021unite}.\\

The contribution of the present work is the introduction of a deep learning energy-conserving interatomic potential for both molecules and materials built on E(3)-equivariant convolutions over geometric tensors that yields state-of-the-art accuracy, outstanding data-efficiency, and can with high fidelity reproduce structural and kinetic properties from molecular dynamics simulations.

\section{Results}

\subsubsection{Equivariance}
The concept of equivariance arises naturally in machine learning of atomistic systems (see e.g. \cite{grisafi2019atomic}): physical properties have well-defined transformation properties under translation, reflection, and rotation of a set of atoms. As a simple example, if a molecule is rotated in space, the vectors of its atomic dipoles or forces also rotate accordingly, via an equivariant transformation. Equivariant neural networks are able to more generally represent tensor properties and tensor operations of physical systems (e.g. vector addition, dot products, and cross products). Equivariant neural networks are guaranteed to preserve the known transformation properties of physical systems under a change of coordinates because they are explicitly constructed from equivariant operations. Formally, a function $f: X \rightarrow Y$ is equivariant with respect to a group $G$ that acts on $X$ and $Y$ if: 

\begin{equation}
    D_Y[g] f(x) = f (D_X[g] x) \quad \forall g \in G, \forall x \in X
\end{equation}
where $D_X[g]$ and $D_Y[g]$ are the representations of the group element $g$ in the vector spaces $X$ and $Y$, respectively. Here, we focus on equivariance with respect to E(3), i.e. the group of rotations, reflections, and translations in 3D space.

\subsubsection{Neural Equivariant Interatomic Potentials}
Given a set of atoms (a molecule or a material), we aim to find a mapping from atomic positions $\{\vec{r}_i\}$ and chemical species $\{Z_i\}$ to the total potential energy $E_{pot}$ and the forces acting on the atoms $\{\vec{F}_i\}$. Following previous work \cite{behler2007generalized}, this total potential energy is obtained as a sum of atomic potential energies. Forces are then obtained as the gradients of this predicted total potential energy with respect to the atomic positions (thereby guaranteeing energy conservation): 
\begin{equation}
    E_{pot} = \sum_{i \in N_{atoms}} E_{i, atomic}
\label{pot_e_sum}
\end{equation}
\begin{equation}
    \vec{F}_i = - \nabla_i E_{pot}
\label{eq:forces_gradient}
\end{equation}
The atomic local energies $E_{i, atomic}$ are the scalar node attributes predicted by the graph neural network. Even though the output of NequIP is the predicted potential energy $E_{pot}$, which is invariant under translations, reflection, and rotations, the network contains \textit{internal features} that are geometric tensors which are equivariant to rotation and reflection. This constitutes the core difference between NequIP and existing scalar-valued invariant GNN-IPs.\\

A series of methods has been introduced to realize rotationally equivariant neural networks \cite{thomas2018tensor, weiler20183d, kondor2018n,  kondor2018clebsch, gnnff}. Here, we build on the layers introduced in Tensor-Field Networks (TFN) \cite{thomas2018tensor}, primitives for which are implemented in \texttt{e3nn} \cite{mario_geiger_2021_4735637}, which enable the construction of neural networks that exhibit equivariance to translation, parity, and rotation. Every atom in NequIP is associated with features comprised of tensors of different orders: scalars, vectors, and higher-order tensors. Formally, the feature vectors are geometric objects that comprise a direct sum of irreducible representations of the O(3) symmetry group. The feature vectors $V_{acm}^{(l,p)}$ are indexed by keys $l,p$, where the ``rotation order'' $l=0, 1,2,...$ is a non-negative integer and parity is one of $p \in (1, -1)$ which together label the irreducible representations of O(3). The indices $a$, $c$, $m$, correspond to the atoms, the channels (elements of the feature vector), and the representation index which takes values $m \in [-l, l]$, respectively. The convolutions that operate on these geometric objects are equivariant functions instead of invariant ones, i.e. if a feature at layer $k$ is transformed under a rotation or parity transformation, then the output of the convolution from layer $k \rightarrow k+1$ is transformed accordingly. 
\\

Convolution operations are naturally translation invariant, since their filters act on relative interatomic distance vectors. Moreover, they are permutation invariant since the sum over contributions from different atoms is invariant to permutations of those atoms. Note that while atomic features are equivariant to permutation of atom indices, globally, the total potential energy of the system is invariant to permutation. To achieve rotation equivariance, the convolution filters $F(\vec{r}_{ij})$ are constrained to be products of learnable radial functions and spherical harmonics, which are equivariant under SO(3) \cite{thomas2018tensor}:
 
\begin{equation}
    F(\vec{r}_{ij}) = R(r_{ij}) Y_m^{(l)}(\hat{r}_{ij})  
\end{equation}
where if $\vec{r}_{ij}$ denotes the relative position from central atom $i$ to neighboring atom $j$, $\hat{r}_{ij}$ and $r_{ij}$ are the associated unit vector and interatomic distance, respectively, and $F(\vec{r}_{ij})$ denotes the corresponding convolutional filter. It should be noted that all learnable weights in the filter lie in the rotationally invariant radial function $R(r_{ij})$. This radial function is implemented as a small multi-layer perceptron that operates on interatomic distances expressed in a basis of choice, $R(r_{ij}): \mathds{R} \rightarrow \mathds{R}^{h}$, where $h$ is the feature dimension: 

\begin{equation}
    R(r_{ij}) = W_n\sigma( ... \sigma(W_2\sigma(W_1 B(r_{ij}))))
\end{equation}
where $B(r_{ij})$ is a basis embedding of the interatomic distance, $W_i$ are weight matrices and $\sigma(x)$ denotes the nonlinear activation function, for which we use the SiLU activation function \cite{hendrycks2016gaussian} in our experiments. Radial Bessel functions and a polynomial envelope function $f_{env}$ \cite{klicpera2020directional} are used as the basis for the interatomic distances: 

\begin{equation}
    B(r_{ij})= \frac{2}{r_c} \frac{\sin(\frac{n\pi}{r_c}r_{ij})}{r_{ij}} f_{env}(r_{ij}, r_c)
\end{equation} 
where $r_c$ is a local cutoff radius, restricting interactions to atoms closer than some cutoff distance and $f_{env}$ is the polynomial defined in \cite{klicpera2020directional} with $p=6$ operating on the interatomic distances normalized by the cutoff radius $\frac{r_{ij}}{r_c}$. The use of cutoffs/local atomic environments allows the computational cost of evaluation to scale linearly with the number of atoms. Similar to \cite{klicpera2020directional}, we initialize the Bessel functions with $n = [1, 2, ..., N_{b}]$, where $N_b$ is the number of basis functions, and subsequently optimize $n \pi$ via backpropagation rather than keeping it constant. For systems with periodic boundary conditions, we use neighbor lists as implemented in the ASE code \cite{Hjorth_Larsen_2017} to identify appropriate atomic neighbors.\\

Finally, in the convolution, the input atomic feature tensor and the filter have to again be combined in an equivariant manner, which is achieved via a geometric tensor product that yields an output feature that again is rotationally equivariant. A tensor product of two geometric tensors is computed via contraction with the Clebsch-Gordan coefficients, as outlined in \cite{thomas2018tensor}. A tensor product between an input feature of order $l_i$ and a convolutional filter of order $l_f$ yields irreducible representations of output orders $|l_i - l_f| \leq l_o \leq |l_i + l_f| $. In NequIP, we use a maximum rotation order $l_{\text{max}}$ and discard all tensor product operations that would results in irreducibe representations with $l_o > l_{\text{max}}$. Omitting all higher-order interactions that go beyond the $0 \otimes 0 \rightarrow 0$ interaction will result in a conventional GNN-IP with invariant convolutions over scalar features, similar to e.g. SchNet \cite{schnet_neurips}.\\

The final symmetry the network needs to respect is that of parity: how the tensor transformations when the input is mirrored, i.e. $\vec{x} \rightarrow -\vec{x}$. A tensor has even parity ($p = 1$) if it is invariant to such a transformation; it has odd parity ($p = -1$) if its sign flips under that transformation. Parity equivariance is achieved by only allowing contributions from a filter and an incoming tensor feature with  parities $p_f$ and $p_i$ to contribute to an output feature if the following selection rule is satisfied:

\begin{equation}
    p_o = p_i p_f
\end{equation}

Finally, as outlined in \cite{thomas2018tensor}, a full convolutional layer $\mathcal{L}$ implementing an interaction with filter $f$ acting on an input $i$ producing output $o$: $l_i \otimes l_f  \rightarrow l_o$ is given by: 

\begin{widetext}
\begin{equation}
    \label{eq:convolution}
    \mathcal{L}_{acm_o}^{(l_o, p_o, l_f, l_i, p_f, p_i)}(\vec{r}_a, V_{acm_i}^{(l_i, p_i)}) = \sum_{m_f, m_i} C_{(l_i, m_i)(l_f, m_f)}^{(l_o, m_o)} \sum_{b \in S} \Big( R_c^{(l_f, l_i, p_f, p_i)}(r_{ab}) \Big) Y_{m_f}^{(l_f)}(\hat{r}_{ab}) V_{bcm_i}^{(l_i, p_i)}
\end{equation}
\end{widetext}
where $a$ and $b$ index the central atom of the convolution and the neighboring atom $b \in S$, respectively, and $C$ indicates the Clebsch-Gordan coefficients. Note that the Clebsch-Gordan coefficients do not depend on the parity of the arguments. There can be multiple $\mathcal{L}_{acm_o}^{(l_o, p_o)}$ tensors for a given output rotation order and parity $(l_o, p_o)$ resulting from different combinations of $(l_i, p_i)$ and $(l_f, p_f)$; we take all such possible output tensors with $l_o \leq l_\text{max}$ and concatenate them. We also divide the output of the sum over neighbors by $\sqrt{N}$, where $N$ denotes the average number of neighbors of an atom. To update the atomic features, the model also uses dense layers that are applied in an atom-wise fashion with weights shared across atoms, similar to the self-interaction layers in SchNet \cite{schnet_neurips}. While different weights are used for different rotation orders, the same set of weights is applied for all representation indices $m$ of a given tensor with rotation order $l$ to maintain equivariance. \\ 

\begin{figure*}[!htbp]
  \includegraphics[width=\linewidth]{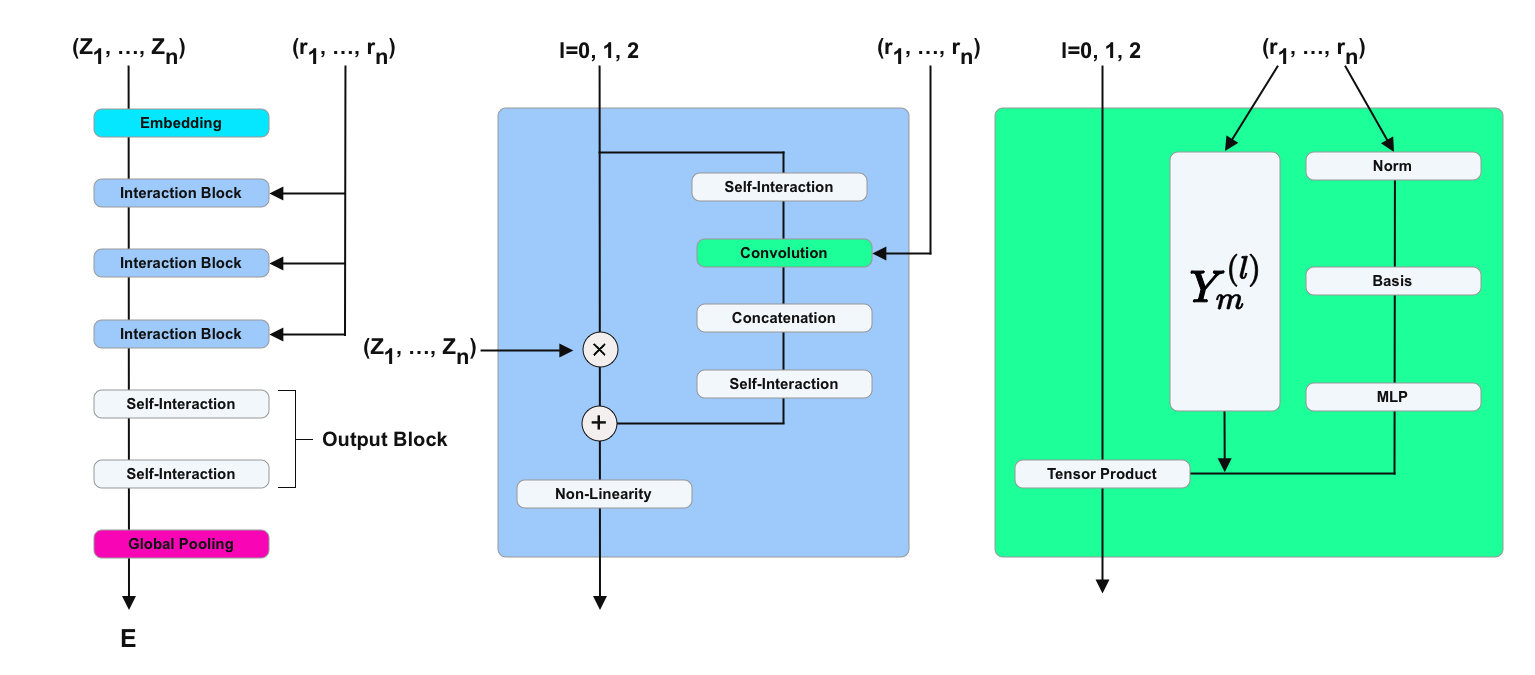}
  \caption{The NequIP network architecture. Left: atomic numbers are embedded into $l=0$ features, which are refined through a series of interaction blocks, creating scalar and higher-order tensor features. An output block then generates atomic energies, which are pooled to give the total predicted energy. Middle: the interaction block, containing the convolution. Right: the convolution combines the radial function $R(r)$ which operates only on interatomic distances with the spherical harmonic projection of the unit vector $\hat{r}_{ij}$ via a tensor product.}
  \label{fig:network}
\end{figure*}

The NequIP network architecture, shown in Figure \ref{fig:network}, is built on an atomic embedding, followed by a series of interaction blocks, and finally an output block: 

\begin{itemize}
    \item \textbf{Embedding}: following SchNet, the initial feature is generated using a trainable embedding that operates on the atomic number $Z_i$ (represented via a one-hot encoding) alone, implemented via a trainable self-interaction layer. 
    \item \textbf{Interaction Block}: interaction blocks encode interactions between neighboring atoms: the core of this block is the convolution function, outlined in equation \ref{eq:convolution}. Features from different tensor product interactions that yield the same rotation and parity pair $(l_o, p_o)$ are mixed by linear atom-wise self-interaction layers. We equip interaction blocks with a ResNet-style update \cite{he2016deep}: $\mathbf{x^{k+1}} = f(\mathbf{x^k}) + \operatorname{Self-Interaction}(\mathbf{x^k})$, where $f$ is the series of self-interaction, convolution, concatenation, and self-interaction. The weights of the $\operatorname{Self-Interaction}$ in the preceding formula are learned separately for each species. Finally, the mixed features are processed by an equivariant SiLU-based gate nonlinearity \cite{hendrycks2016gaussian, weiler20183d} (even and odd scalars are not gated, but instead are processed directly by SiLU and tanh nonlinearities, respectively). 
    
    \item \textbf{Output Block}: the $l=0$ features of the final convolution are passed to an output block, which consists of a set of two atom-wise self-interaction layers.
\end{itemize}

For each atom the final layer outputs a single scalar, which is interpreted as the atomic potential energy. These are then summed to give the total predicted potential energy of the system (Equation \ref{pot_e_sum}). Forces are subsequently obtained as the negative gradient of the predicted total potential energy, thereby ensuring both energy conservation and rotation-equivariant forces (see equation \ref{eq:forces_gradient}).  

\subsection{Experiments}

We validate the proposed method on a diverse series of challenging data sets: first we demonstrate that we improve upon state-of-the-art accuracy on MD-17, a data set of small, organic molecules that is widely used for benchmarking ML-IPs \cite{chmiela2017machine, schutt2017quantum, sgdml, schnet_neurips, klicpera2020directional}. Next, we show that NequIP can accurately learn forces obtained on small molecules at the quantum chemical CCSD(T) level of theory \cite{sgdml}. To broaden the applicability of the method beyond small isolated molecules, we finally explore a series of extended systems with periodic boundary conditions, consisting of both surfaces and bulk materials: water in different phases \cite{deepmd, deepmd_water_paper}, a chemical reaction at a solid/gas interface, an amorphous Lithium Phosphate \cite{dcf}, and a Lithium superionic conductor \cite{gnnff}. Details of the training procedure are provided in the Methods section.

\subsubsection{MD-17 small molecule dynamics}
We first evaluate NequIP on MD-17 \cite{chmiela2017machine, schutt2017quantum, sgdml}, a data set of small organic molecules in which reference values of energy and forces are generated by ab-initio MD simulations with DFT. Recently, a recomputed version of the original MD-17 data with higher numerical accuracy has been released, termed the revised MD-17 data set \cite{christensen2020role} (an example histogram of potential energies and force components can be found in Appendix C). In order to be able to compare results to a wide variety of methods, we benchmark NequIP on both data sets. For training and validation, we use a combined N=1,000 configurations. The mean absolute error in the energies and force components is shown in Tables \ref{tab:md17} and \ref{tab:revmd17}. We compare results using NequIP with those from published leading MLIP models. We find that NequIP significantly outperforms invariant GNN-IPs (such as SchNet \cite{schnet_neurips} and DimeNet \cite{klicpera2020directional}), shallow neural networks (such as ANI \cite{devereux2020extending}), and kernel-based approaches (such as GAP \cite{gaporiginalpaper}, FCHL19/GPR \cite{christensen2020fchl, christensen2020role} and sGDML \cite{sgdml}). Finally, we compare to a series of other methods including ACE \cite{drautz2019atomic}, SpookyNet \cite{unke2021}, and GemNet \cite{klicpera2021gemnet} as well as other equivariant neural networks such as PaiNN \cite{schutt2021equivariant}, NewtonNet \cite{haghighatlari2021newtonnet}, and UNiTE \cite{qiao2021unite}. Again, it should be stressed that PaiNN and NewtonNet are $l_{max}=1$-only versions of equivariant networks. The results for ACE, GAP, and ANI on the revised MD-17 data set are those reported in \cite{kovacs2021linear}. Importantly, we train and test separate NequIP models on both the original and the revised MD-17 data set, and find that NequIP obtains significantly lower energy errors on the revised data set, while the force accuracy is similar on the two data sets. In line with previous work \cite{christensen2020role}, this suggests that the noise floor on the original MD-17 data is higher on the energies and that only the results on the revised MD-17 data set should be used for comparing different methods.\\

Remarkably, we find that NequIP outperforms all other methods. The consistent improvements in accuracy compared to sGDML and FCHL19/GPR are particularly surprising, as these are based on kernel methods, which typically obtain better performance than deep neural networks on small training sets. We run a convergence scan on the rotation order $l \in \{0, 1, 2, 3\}$ and find that increasing the tensor rank beyond $l=1$ gives a consistent improvement. The significant improvement from $l=0$ to $l=1$ highlights the crucial role of equivariance in obtaining improved accuracy on this task. 

\begin{table*}[!htbp]
\centering
\resizebox{\textwidth}{!}{\begin{tabular}{llccccccccc}
\hline \hline
Molecule & & SchNet & DimeNet & sGDML & PaiNN & SpookyNet & GemNet-(T/Q) & NewtonNet  & UNiTE & NequIP (l=3) \\
\hline
\multirow{2}{*}{Aspirin}  & \textit{Energy} & 16.0& 8.8 & 8.2 & 6.9 & 6.5&-  & 7.3 & - & \textbf{5.7}\\ 
                          & \textit{Forces} & 58.5  &21.6  & 29.5& 14.7& 11.2 & 9.4&  15.1 & \textbf{6.8}& 8.2 \\ \hline

\multirow{2}{*}{Ethanol}  & \textit{Energy} & 3.5& 2.8&3.0 & 2.7 & 2.3 & - & 2.6 & - & \textbf{2.2} \\ 
                          & \textit{Forces} & 16.9 &10.0  & 14.3 &9.7 & 4.1&  \textbf{3.7} & 9.1 & 4.0 & 3.8 \\ \hline
                          
\multirow{2}{*}{Malonaldehyde}  & \textit{Energy} & 5.6& 4.5 &4.3 & 3.9 & 3.4 &-  & 4.2& -  & \textbf{3.3} \\ 
                          & \textit{Forces} &28.6 & 16.6& 17.8 & 13.8&7.2 & 6.7& 14.0& 6.9 & \textbf{5.8}  \\ \hline
                          
\multirow{2}{*}{Naphthalene}  & \textit{Energy} & 6.9 &5.3 & 5.2& 5.0 &5.0  & - & 5.1&  - & \textbf{4.9}\\ 
                          & \textit{Forces} & 25.2& 9.3& 4.8  & 3.3& 3.9& 2.2&  3.6 & 2.8 & \textbf{1.6} \\ \hline

\multirow{2}{*}{Salicylic acid}  & \textit{Energy} & 8.7& 5.8 & 5.2 & 4.9 & 4.9 & - & 5.0& -  &  \textbf{4.6}\\ 
                          & \textit{Forces} & 36.9& 16.2  & 12.1 & 8.5 &7.8 & 5.4 & 8.5& 4.2 & \textbf{3.9} \\ \hline
                          
\multirow{2}{*}{Toluene}  & \textit{Energy} & 5.2 & 4.4 &4.3 & 4.1 &4.1 & - & 4.1&  - &  \textbf{4.0}\\ 
                          & \textit{Forces} & 24.7 & 9.4 & 6.1 & 4.1 & 3.8& 2.6 & 3.8 & 3.1 & \textbf{2.0}  \\ \hline
                          
\multirow{2}{*}{Uracil}  & \textit{Energy} & 6.1 & 5.0 &4.8 &\textbf{ 4.5} & 4.6& - & 4.6&  - & \textbf{4.5} \\ 
                          & \textit{Forces} & 24.3& 13.1 &10.4 & 6.0 &5.2 & 4.2 & 6.5 & 4.2 & \textbf{3.3}\\ \hline
  \hline \hline
    \end{tabular}}
    \caption{Energy and Force MAE for molecules on the \textbf{original MD-17 data set}, reported in units of [meV] and [meV/{\AA}], respectively, and a training budget of 1,000 reference configurations. For GemNet, the best result out of the T/Q versions is presented and for PaiNN the best between force-only and joint force and energy training. For UNiTE, we compare to the "direct-learning" results reported in \cite{qiao2021unite}. Best results are marked in \textbf{bold}.}
    \label{tab:md17}
\end{table*}

\begin{table*}[!htbp]
\centering
\resizebox{\textwidth}{!}{\begin{tabular}{llcccccccccc}
\hline \hline
Molecule & & FCHL19 & UNiTE & GAP & ANI & ACE & GemNet-(T/Q) & NequIP (l=0) & NequIP (l=1) & NequIP (l=2)  & NequIP (l=3) \\
\hline
\multirow{2}{*}{Aspirin}  & \textit{Energy} & 6.2 & 2.4 & 17.7 & 16.6 & 6.1 & - & 25.2 & 3.8  & 2.4 & \textbf{2.3}  \\ 
                          & \textit{Forces} & 20.9 & \textbf{7.6}& 44.9 & 40.6 & 17.9 & 9.5 & 41.9 & 12.9 & 8.7 & 8.5 \\ \hline

\multirow{2}{*}{Azobenzene}  & \textit{Energy} & 2.8 & 1.1 & 8.5 & 15.9 & 3.6 & - & 20.3 & 1.1 & 0.8 & \textbf{0.7} \\ 
                          & \textit{Forces} & 10.8 & 4.2 & 24.5 & 35.4 & 10.9 & - & 42.3 & 5.6 & 4.2 &\textbf{ 3.6} \\ \hline
                          
\multirow{2}{*}{Benzene}  & \textit{Energy} & 0.3 & 0.07 & 0.75 & 3.3 & \textbf{ 0.04} & - & 3.2 & 0.09 & 0.06&  \textbf{0.04}\\ 
                          & \textit{Forces} & 2.6 & 0.73 & 6.0 & 10.0 & 0.5 & 0.5 & 10.3 & 0.4  & 0.4 &  \textbf{0.3} \\ \hline
                          
\multirow{2}{*}{Ethanol}  & \textit{Energy} & 0.9 & 0.62 & 3.5 & 2.5 & 1.2 & - & 2.0 & 1.0 & 0.5 &  \textbf{0.4 } \\ 
                          & \textit{Forces} & 6.2 & 3.7 & 18.1 & 13.4 & 7.3 & 3.6 & 13.7 & 7.6  &4.2  & \textbf{3.4}   \\ \hline
                          
\multirow{2}{*}{Malonaldehyde}  & \textit{Energy} & 1.5 & 1.1 & 4.8 & 4.6 & 1.7 & - &  4.4 & 1.6 & 0.9 & \textbf{0.8}  \\ 
                          & \textit{Forces} & 10.2 & 6.6 & 26.4 & 24.5 & 11.1 & 6.6 & 23.4 &  10.4 & 6.0  & \textbf{ 5.2} \\ \hline
                          
\multirow{2}{*}{Naphthalene}  & \textit{Energy} & 1.2 & 0.46 & 3.8 & 11.3 & 0.9 & - & 14.7 & 0.4  & 0.3 & \textbf{ 0.2} \\ 
                          & \textit{Forces} & 6.5 & 2.6 & 16.5 & 29.2 & 5.1 & 1.9 & 20.1 & 2.0 & 1.3 &  \textbf{1.2} \\ \hline

\multirow{2}{*}{Paracetamol}  & \textit{Energy} & 2.9 &  1.9 & 8.5 & 11.5 & 4.0  & - & 17.5 & 2.1 & \textbf{1.4} & \textbf{1.4}  \\ 
                          & \textit{Forces} & 12.2 & 7.1  & 28.9 & 30.4 & 12.7 & - & 37.6 & 10.8 &\textbf{ 6.9} & \textbf{ 6.9} \\ \hline

\multirow{2}{*}{Salicylic acid}  & \textit{Energy} & 1.8 & 0.73& 5.6 & 9.2  & 1.8 &  - & 11.4 & 1.0 & 0.8  &\textbf{ 0.7 } \\ 
                          & \textit{Forces} & 9.5 & 3.8 & 24.7 & 29.7 & 9.3 & 5.3 & 28.7 & 5.7 & 4.2 & \textbf{4.0 }\\ \hline
                          
\multirow{2}{*}{Toluene}  & \textit{Energy} & 1.6 & 0.45 & 4.0 & 7.7 & 1.1 & - & 9.7 & 0.5  & \textbf{0.3}  &  \textbf{0.3} \\ 
                          & \textit{Forces} & 8.8 & 2.5 & 17.8 & 24.3 & 6.5 & 2.2 & 27.2 & 2.7 & 1.8 & \textbf{1.6} \\ \hline
                          
\multirow{2}{*}{Uracil}  & \textit{Energy} & 0.6 & 0.58 & 3.0 & 5.1 & 1.1 & - & 10.0 & 0.6&\textbf{ 0.4} &\textbf{ 0.4 } \\ 
                          & \textit{Forces} & 4.2 & 3.8 & 17.6 & 21.4 & 6.6 & 3.8 & 25.8 & 4.1 &\textbf{ 3.0} & 3.2  \\ \hline
  \hline \hline
    \end{tabular}}
    \caption{Energy and Force MAE for molecules on the \textbf{revised MD-17 data set}, reported in units of [meV] and [meV/{\AA}], respectively, and a training budget of 1,000 reference configurations. For GemNet, the best result out of the T/Q versions is presented. For FCHL19, the best results between energy-only, force-only and joint force and energy training are presented. For UNiTE, we compare to the "direct-learning" results reported in \cite{qiao2021unite}. Best results are marked in \textbf{bold}. }
    \label{tab:revmd17}
\end{table*}

\subsubsection{Force training at quantum chemical accuracy}
The ability to achieve high accuracy on a comparatively small data set facilitates easier development of Machine Learining Interatomic Potentials on expensive high-order \textit{ab-initio} quantum chemical methods, such as e.g. the coupled cluster method CCSD(T). However, the high computational cost of CCSD(T) has thus far hindered the use of reference data structures at this level of theory, prohibited by the need for large data sets that are required by available NN-IPs. Leveraging the high data efficiency of NequIP, we evaluate it on a set of molecules computed at quantum chemical accuracy (aspirin at CCSD, all others at CCSD(T)) \cite{sgdml} and compare the results to those reported for sGDML \cite{sgdml} and GemNet \cite{klicpera2021gemnet} in table \ref{tab:ccsdt}.

\begin{table*}[!htbp]
\centering
\begin{tabular}{llccc}
\hline \hline
Molecule & & sGDML & GemNet-(T/Q) & NequIP (l=3) \\
\hline
\multirow{2}{*}{Aspirin}  & \textit{Energy} & 6.9 & - & \textbf{2.0} \\ 
                          & \textit{Forces} & 33.0 & 10.3 & \textbf{ 8.3} \\ \hline

\multirow{2}{*}{Benzene}  & \textit{Energy} & 0.17 & - & \textbf{0.05} \\ 
                          & \textit{Forces} & 1.7 & 0.7 & \textbf{0.26}\\ \hline
                          
\multirow{2}{*}{Ethanol}  & \textit{Energy} & 2.2 & - & \textbf{0.36} \\ 
                          & \textit{Forces} & 15.2 & 3.1 & \textbf{3.0}  \\ \hline
                          
\multirow{2}{*}{Malonaldehyde}  & \textit{Energy} &2.6  & - & \textbf{0.72} \\ 
                          & \textit{Forces} & 16.0 & 5.9 & \textbf{4.5}  \\ \hline
                          
\multirow{2}{*}{Toluene}  & \textit{Energy} & 1.3 & - & \textbf{0.27} \\ 
                          & \textit{Forces} & 9.1 & 2.7 & \textbf{1.7} \\ \hline
  \hline \hline
    \end{tabular}
    \caption{Energy and Force MAE for molecules at CCSD/CCSD(T) accuracy, reported in units of [meV] and [meV/{\AA}], respectively, and a training budget of 1,000 reference configurations. For GemNet, the best result out of the T/Q versions is presented.}
    \label{tab:ccsdt}
\end{table*}

\begin{table*}[!htbp]
\centering
\begin{tabular}{llccccc}
\hline \hline
System & &   NequIP, a) & NequIP, b) & NequIP, c)  & DeepMD  \\
\hline
\multirow{2}{*}{Liquid Water}  
                          & \textit{Energy} & -  & 1.6  & 1.7 & 1.0 \\
                          & \textit{Forces} &  12.5 & 51.4 & 12.2 & 40.4 \\ \hline
                          
\multirow{2}{*}{Ice Ih (b)}   
                          & \textit{Energy} & -  & 2.5 & 4.3 & 0.7 \\
                          & \textit{Forces} &  10.8  & 57.8  & 10.4 & 43.3 \\ \hline
                          
\multirow{2}{*}{Ice Ih (c)}  
                         & \textit{Energy} & -  &  3.9 & 10.2 & 0.7 \\
                          & \textit{Forces} &  12.5 & 29.1 & 12.2 & 26.8 \\ \hline
                          
\multirow{2}{*}{Ice Ih (d)} 
                          & \textit{Energy} &  - &  2.6  & 12.7 & 0.8 \\
                          & \textit{Forces} &  10.3 &  24.1 & 10.1 & 25.4\\ \hline

  \hline \hline
    \end{tabular}
    \caption{RMSE of energies and forces on liquid water and the three ices in units of [meV/molecule] and [meV/A], with energy errors normalized by the number of molecules in the system. Note that the NequIP models were trained on $<0.1\%$ of the training data of DeepMD. NequIP model a) refers to loss function weighting $\lambda_F=1, \lambda_E=0$, model b) to $\lambda_F=100, \lambda_E=1$, and model c) to $\lambda_F=100,000, \lambda_E=1$.}
    \label{tab:deepmd}
\end{table*}

\begin{table}[!htbp]
\centering
\begin{tabular}{llc}
\hline \hline
Element & & MAE \\
\hline
C & \textit{Forces} & 19.9  \\ \hline

O  & \textit{Forces} & 73.1  \\ \hline
                    
H & \textit{Forces} &  13.0 \\ \hline
                    
Cu & \textit{Forces} &  47.6 \\ \hline \hline

\multirow{2}{*}{Total}  & \textit{Energy} & 0.50   \\ 
                    & \textit{Forces} &  38.4 \\ \hline 
  \hline \hline
    \end{tabular}
    \caption{Nequip MAE of energies and force components for Formate on Cu system, per-element basis. The training set consists of 2,500 structures, units are [meV/atom] and [meV/A], respectively.}
    \label{table:fcu}
\end{table}

\subsubsection{Liquid Water and Ice Dynamics}

To demonstrate the applicability of NequIP beyond small molecules, we evaluate the method on a series of extended systems with periodic boundary conditions. As a first example we use a joint data set consisting of liquid water and three ice structures \cite{deepmd, deepmd_water_paper} computed at the PBE0-TS level of theory. This data set \cite{deepmd} contains: a) liquid water, P=1bar, T=300K, computed via path-integral AIMD, b) ice Ih, P=1bar, T=273K, computed via path-integral AIMD c) ice Ih, P=1bar, T=330K, computed via classical AIMD d) ice Ih, P=2.13 kbar, T=238K, computed via classical AIMD. A DeepMD NN-IP model was previously trained \cite{deepmd} for water and ice using a joint training set containing 133,500 reference calculations of these four systems. To assess data efficiency of the NequIP architecture, we similarly train a model jointly on all four parts of the data set, but using only 133 structures for training, i.e. 1000x fewer data. The 133 structures were sampled randomly following a uniform distribution from the full data set available online which consists of water and ice structures and is made up of a total of 140,000 frames, coming from the same MD trajectories that were used in the earlier work \cite{deepmd}. Table \ref{tab:deepmd} compares the energy and force errors of NequIP trained on the 133 structures vs DeepMD trained on 133,500 structures.  We find that with 1000x fewer training data NequIP significantly outperforms DeepMD on all four parts of the data set in the error on the force components. We note that there are $3N$ force components for each training frame but only one energy target. Consequently, one would except that on energies the much larger training set used for DeepMD would results in an even stronger difference. We find that while this is indeed the case, the NequIP results on the liquid phase are surprisingly competitive. Finally, we report results using three different weightings of energies and forces in the loss function and see that increasing the energy weighting results in significantly improved energy errors at the cost of a small increase in force error. We note that the version of DeepMD published in \cite{deepmd} is not smooth, and a smooth version has since been proposed \cite{zhang2018end}. However, \cite{zhang2018end} does not report results on the water/ice systems. It would be of interest to investigate the performance of the smooth DeepMD version as a function of training set size.

\subsubsection{Heterogeneous catalysis of formate dehydrogenation}
\begin{figure}[!htbp]
  \includegraphics[width=\linewidth]{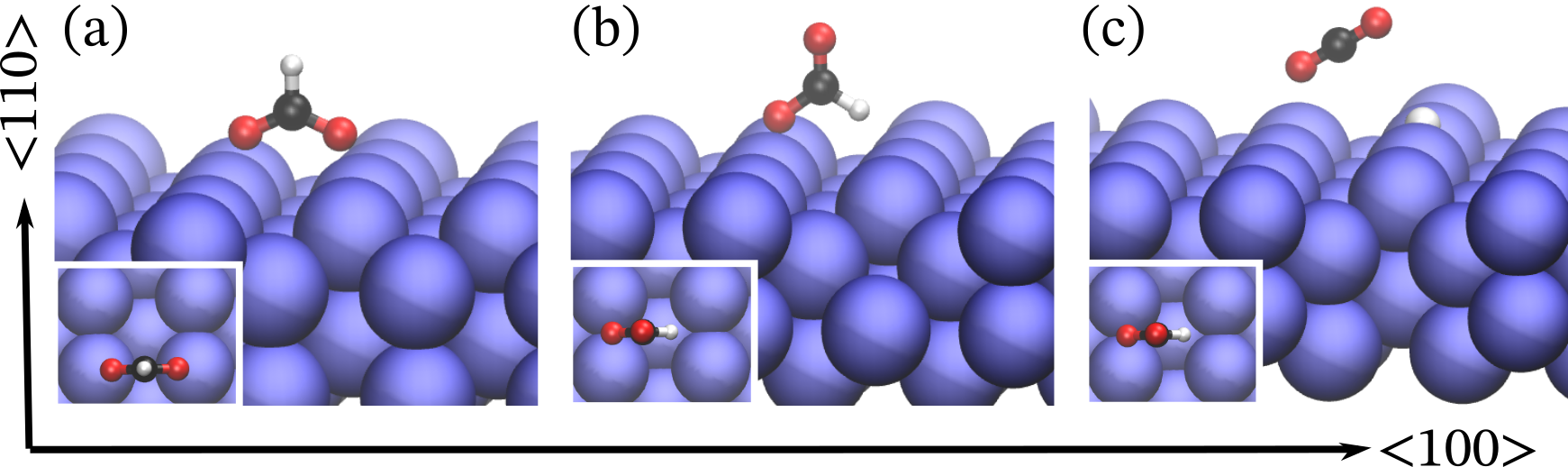}
  \caption{Perspective view of atomic configurations of (a) bidentate HCOO (b) monodentate HCOO and (c) CO$_2$ and a hydrogen adatom on a Cu(110) surface. The blue, red, black, and white spheres represent Cu, O, C, and H atoms, respectively. The subset shown in each subplot is the corresponding top view along the $<110>$ orientation.}
  \label{fig:fcu}
\end{figure}

Next, we apply NequIP to a catalytic surface reaction. In particular, we investigate the dynamics of formate undergoing dehydrogenation decomposition ($\mathrm{HCOO}^* \rightarrow \mathrm{H}^* + \mathrm{CO}_2$) on a Cu $<110>$ surface (see Figure \ref{fig:fcu}). This system is highly heterogeneous: it has both metallic and covalent types of bonding as well as charge transfer between the metal and the molecule, making it a particularly challenging test system. Different states of the molecule also lead to dissimilar C-O bond lengths \cite{Sim_Gardner_King_1996,Wang_Morikawa_Matsumoto_Nakamura_2006}. Training structures consist of 48 Cu atoms and 4 atoms of the molecule (HCOO* or CO$_2$+H*). The MAE of the predicted forces using a NequIP model trained on 2,500 structures is shown in Table \ref{table:fcu}, demonstrating that NequIP is able to accurately model the interatomic forces for this complex reactive system. A more detailed analysis of the resulting dynamics will be the subject of a separate study.\\

\subsubsection{Lithium Phosphate Amorphous Glass Formation}

\begin{figure}[!htbp]
  \includegraphics[width=.8\linewidth]{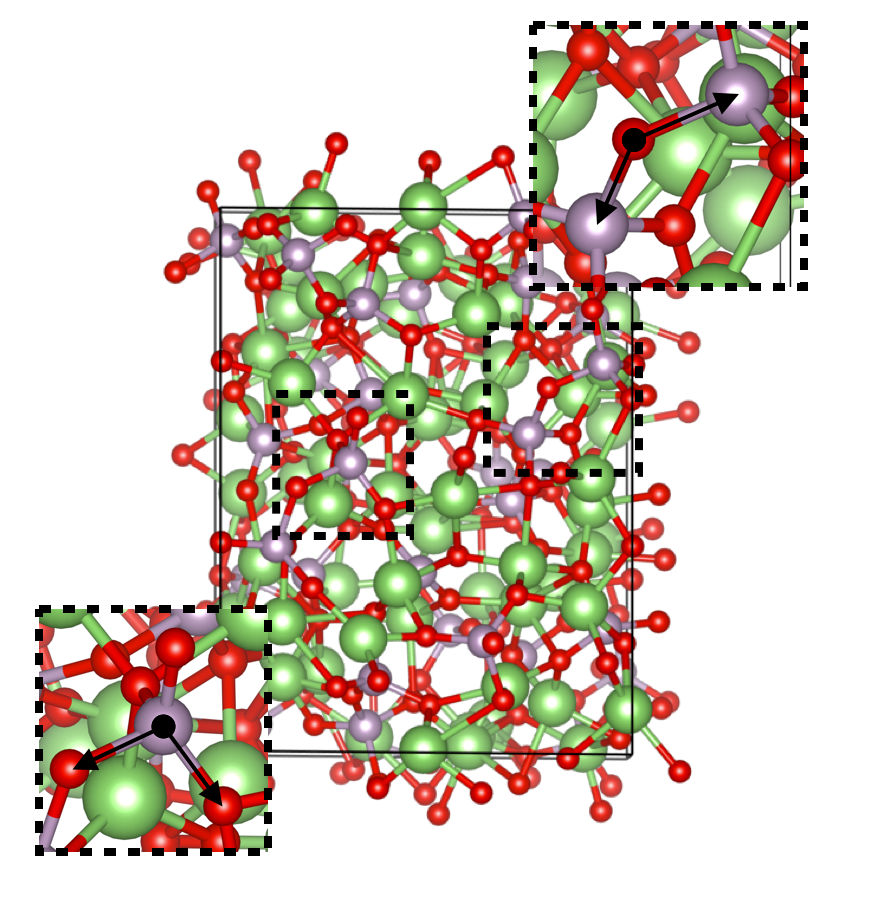}
  \caption{Quenched glass structure of Li\textsubscript{4}P\textsubscript{2}O\textsubscript{7}. The insets show  the P-O-O tetrahedral bond angle (bottom left) as well as the O-P-P bridging angle between corner-sharing phosphate tetrahedra (top right).}
  \label{fig:lipo}
\end{figure}

\begin{figure*}[!htb]
\minipage{0.32\textwidth}
  \includegraphics[width=\linewidth]{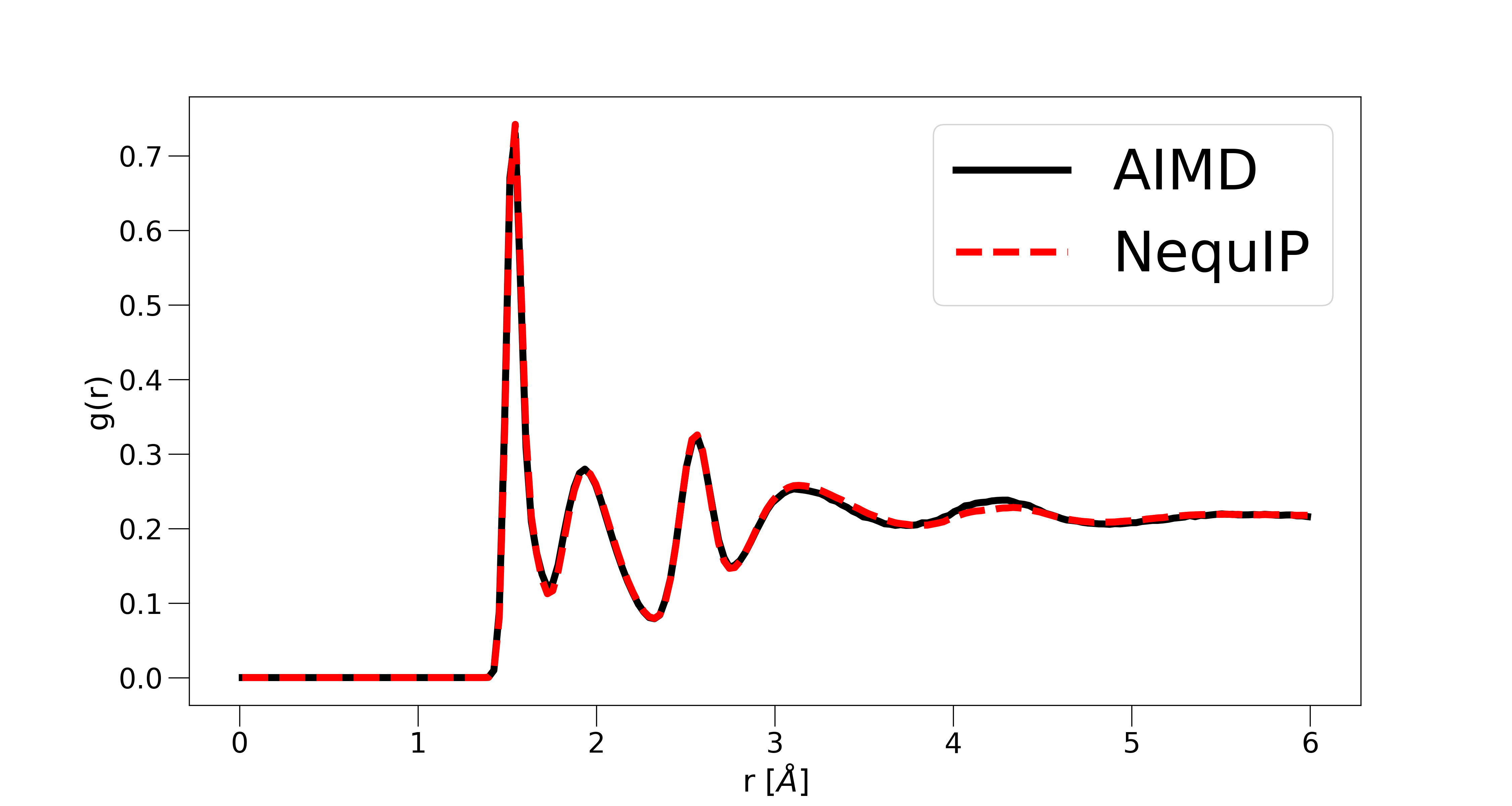}
\endminipage\hfill
\minipage{0.32\textwidth}
  \includegraphics[width=\linewidth]{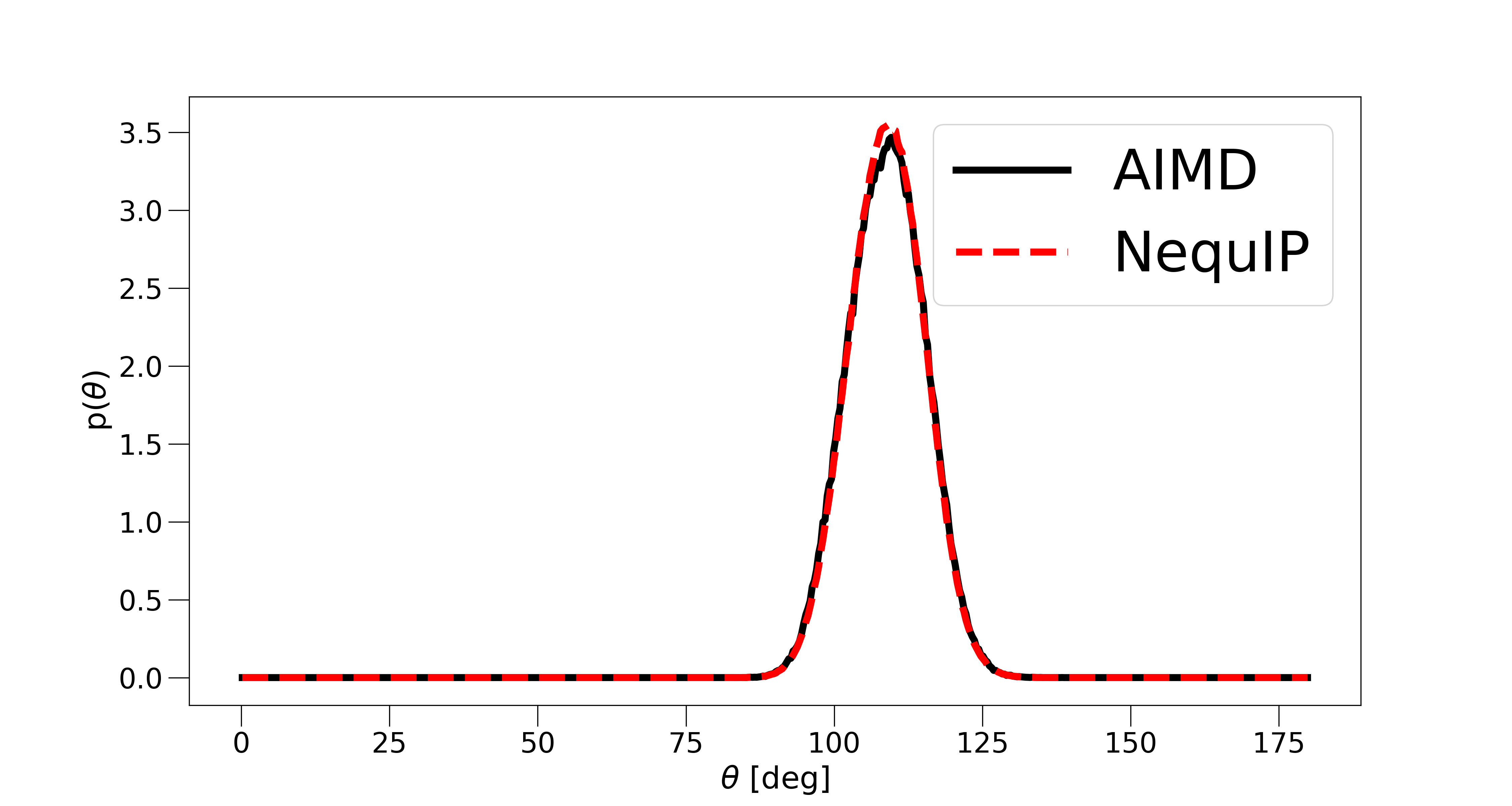}
\endminipage\hfill
\minipage{0.32\textwidth}%
  \includegraphics[width=\linewidth]{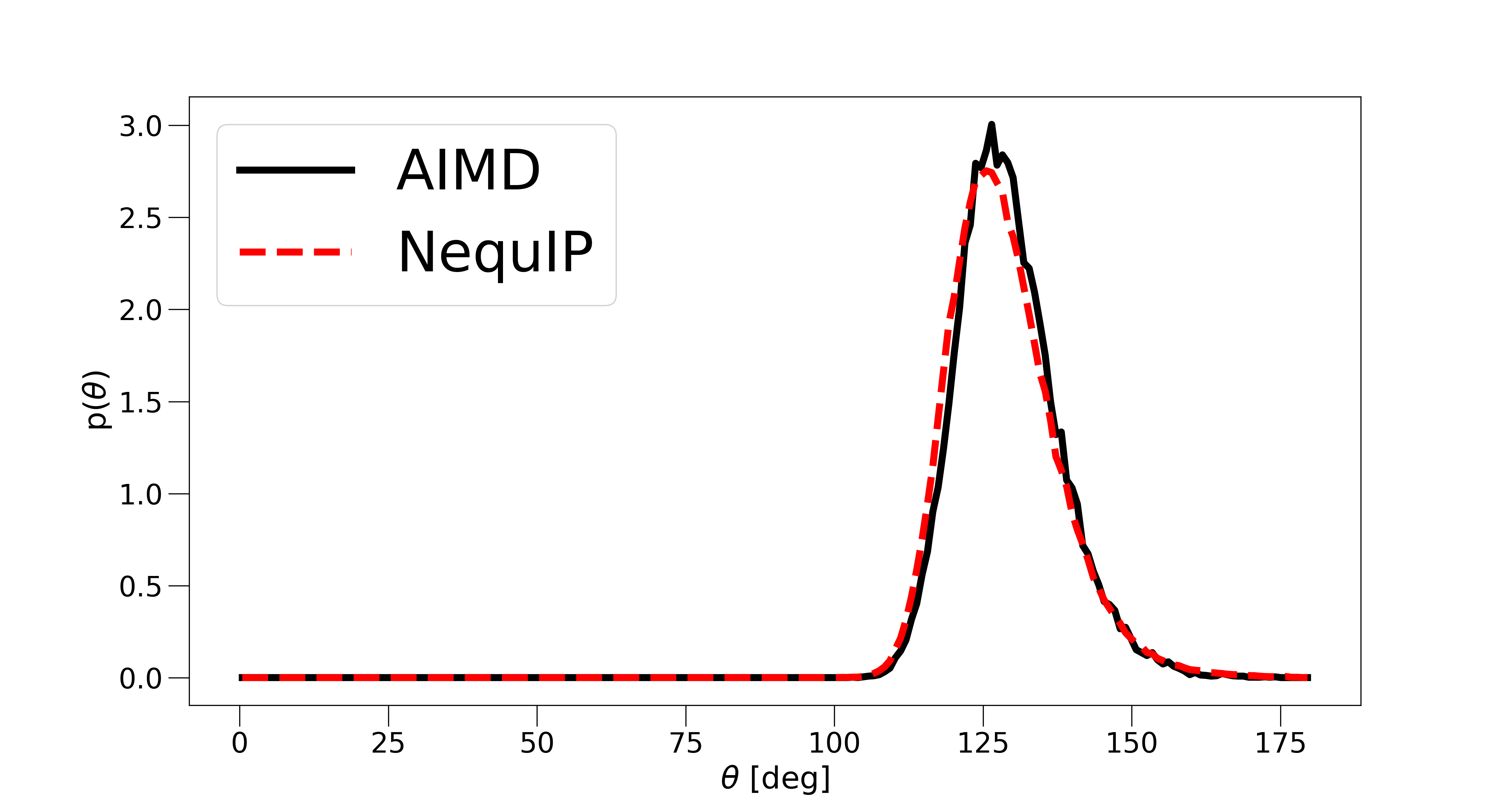}
\endminipage
\caption{Left: Radial Distribution Function, middle:  Angular Distribution Function, tetrahedral bond angle, right: Angular Distribution Function, bridging oxygen. All are defined as probability density functions; NequIP results are averaged over 10 runs with different initial velocities.}
\label{fig:structure}
\end{figure*}

To examine the ability of the model to capture dynamical properties, we demonstrate that NequIP can describe structural dynamics in amorphous lithium phosphate with composition Li\textsubscript{4}P\textsubscript{2}O\textsubscript{7}. This material is a member of the promising family of solid electrolytes for Li-metal batteries \cite{LiPON,LiSiPON, dcf}, with non-trivial Li-ion transport and phase transformation behaviors. The data set consists of two 50ps-long AIMD simulations: one of the molten structure at T=3000 K and another of a quenched glass structure at T=600 K. We train NequIP on a subset of 1,000 structures from the molten trajectory. Table \ref{tab:lps} shows the error in the force components on both the test set from the AIMD molten trajectory and the full AIMD quenched glass trajectory. To then evaluate the physical fidelity of the trained model, we use it to run a set of ten MD simulations of length 50 ps at T=600 K in the NVT ensemble and compare the total radial distribution function (RDF) without element distinction as well as the angular distribution functions (ADF) of the P-O-O (P central atom) and  O-P-P (O central atom) angles averaged over ten runs to the \textit{ab-inito} trajectory at the same temperature. The P-O-O angle corresponds to the tetrahedral bond angle, while the O-P-P corresponds to a bridging angle between corner-sharing phosphate tetrahedra (Figure \ref{fig:lipo}). Figure \ref{fig:structure} shows that NequIP can accurately reproduce the RDF and the two ADFs, in comparison with AIMD, after training on only 1,000 structures. This demonstrates that the model generates the glass state and recovers its dynamics and structure almost perfectly, despite having seen only the high-temperature molten training data. We also include results from a longer NequIP-driven MD simulation of 500 ps, which can be found in Appendix A.

\subsubsection{Lithium Thiophosphate Superionic Transport}
\begin{figure}[!htbp]

  \includegraphics[width=\linewidth]{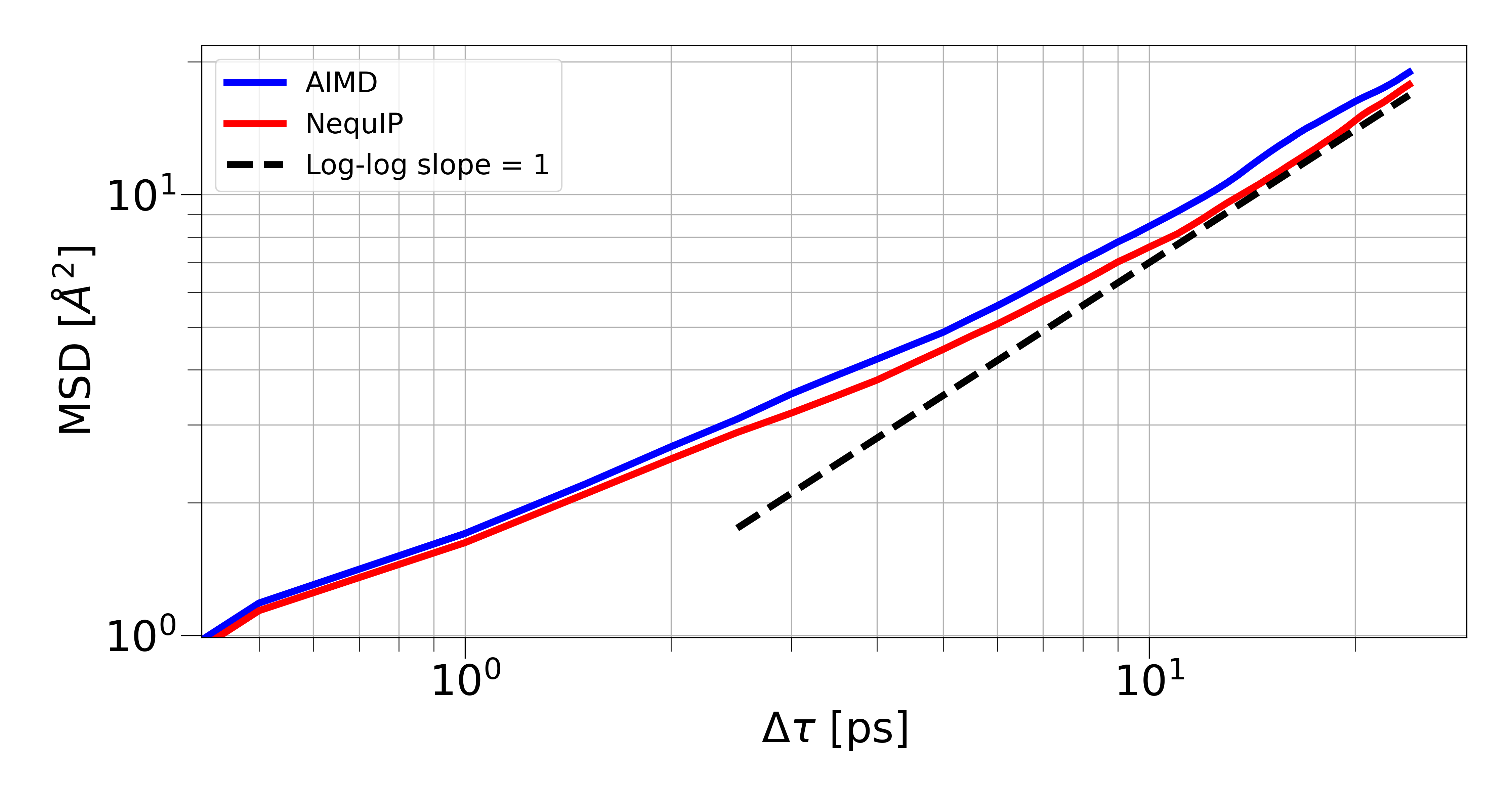}
  \caption{Comparison of the Lithium mean square displacement of AIMD and an example NequIP trajectory.}
  \label{fig:lps_msd}
\end{figure}

To show that NequIP can model kinetic transport properties from small training sets at high accuracy, we study Li-ion diffusivity in LiPS (Li\textsubscript{6.75}P\textsubscript{3}S\textsubscript{11}) a crystalline superionic Li conductor, consisting of a simulation cell of 83 atoms \cite{gnnff}. MD is widely used to study diffusion; training a ML-IP to the accuracy required to predict kinetic properties, however, has in the past required large training set sizes (\cite{li2017study} e.g. uses a data set of 30,874 structures to study Li diffusion in Li\textsubscript{3}PO\textsubscript{4}). Here we demonstrate that not only does NequIP obtain small errors in the energies and force components, but it also accurately predicts the diffusivity after training on a data set obtained from an AIMD simulation. Again, we find that very small training sets lead to highly accurate models, as shown in Table \ref{tab:lps} for training set sizes of 10, 100, 1,000 and 2,500 structures. We run a series of MD simulations with the NequIP potential trained on 2,500 structures in the NVT ensemble at the same temperature as the AIMD simulation for a total simulation time of 50 ps and a time step of 0.25 fs, which we found advantageous for the reliability and stability of long simulations. We measure the Li diffusivity in these NequIP-driven MD simulations (computed via the slope of the mean square displacement) started from different initial velocities, randomly sampled from a Maxwell-Boltzmann distribution. We find a mean diffusivity of $1.25\operatorname{x}10^{-5} \operatorname{cm^2/s}$, in excellent agreement with the diffusivity of $1.37 \operatorname{x}10^{-5} \operatorname{cm^2/s}$ computed from AIMD, thus achieving a relative error of as little as $9\%$. Figure \ref{fig:lps_msd} shows the mean square displacements of Li for an example run of NequIP in comparison to AIMD.

\begin{table*}[!htbp]
\centering
\begin{tabular}{lllcc}
\hline \hline
System & Data set size &  & MAE & RMSE \\
\hline
\multirow{2}{*}{LiPS}  & \multirow{2}{*}{10}  & \textit{Energy} & 2.03 &  2.54 \\ 
                          & & \textit{Forces}  & 109.9 & 142.0 \\ \hline

\multirow{2}{*}{LiPS}  & \multirow{2}{*}{100}  & \textit{Energy} & 0.44 & 0.56  \\ 
                          & & \textit{Forces}  & 28.4 & 36.8 \\ \hline
                          
\multirow{2}{*}{LiPS}  & \multirow{2}{*}{1,000}  & \textit{Energy} &  0.12 & .15 \\ 
                          & & \textit{Forces}  & 8.3 & 11.2 \\ \hline
                          
\multirow{2}{*}{LiPS}  & \multirow{2}{*}{2,500}  & \textit{Energy} & 0.08 & 0.10 \\ 
                          & & \textit{Forces}  & 4.9 & 6.6 \\ \hline \hline

\multirow{2}{*}{Li\textsubscript{4}P\textsubscript{2}O\textsubscript{7}, melt}  & \multirow{2}{*}{1,000}  & \textit{Energy} & 0.4 & 0.8 \\ 
                          & & \textit{Forces}  & 38.6 & 62.7 \\ \hline
                          
\multirow{2}{*}{Li\textsubscript{4}P\textsubscript{2}O\textsubscript{7}, quench}  & \multirow{2}{*}{1,000}  & \textit{Energy} & 0.5 & 0.5 \\ 
                          & & \textit{Forces}  & 24.8  & 38.1\\ \hline
                                       
  \hline \hline
    \end{tabular}
    \caption{NequIP E/F MAE/RMSE for LiPS and Li\textsubscript{4}P\textsubscript{2}O\textsubscript{7} for different data set sizes in units of [meV/A] and [meV/atom]. The model for Li\textsubscript{4}P\textsubscript{2}O\textsubscript{7} was trained exclusively on structures from the melted trajectory. The reported test errors for the melt are computed on the remaining set of structures from the full melt trajectory;  errors for the quench are computed on the full quench trajectory. }
    \label{tab:lps}
\end{table*}

\subsection{Data Efficiency}

In the above experiments, NequIP exhibits exceptionally high data efficiency. It is interesting to consider the reasons for such high performance and verify that it is connected to the equivariant nature of the model. First, it is important to note that each training configuration contains multiple labels: in particular, for a training set of $M$ first-principles calculations with structures consisting of $N$ atoms, the energies and force components together give a total of  $M (3N + 1)$ labels. In order to gain insight into the reasons behind increased accuracy and data efficiency, we perform a series of experiments with the goal of isolating the effect of using equivariant convolutions. In particular, we run a set of experiments in which we explicitly turn on or off interactions of higher order than $l=0$. This defines two settings: first, we train the network with the full set of tensor features up to a given order $l$ and the corresponding equivariant interactions. Second, we turn off all interactions involving $l>0$, making the network a conventional invariant GNN-IP, involving only invariant convolutions over scalar features in a SchNet-style fashion.\\

\begin{figure*}[!htbp]
  \includegraphics[width=\linewidth]{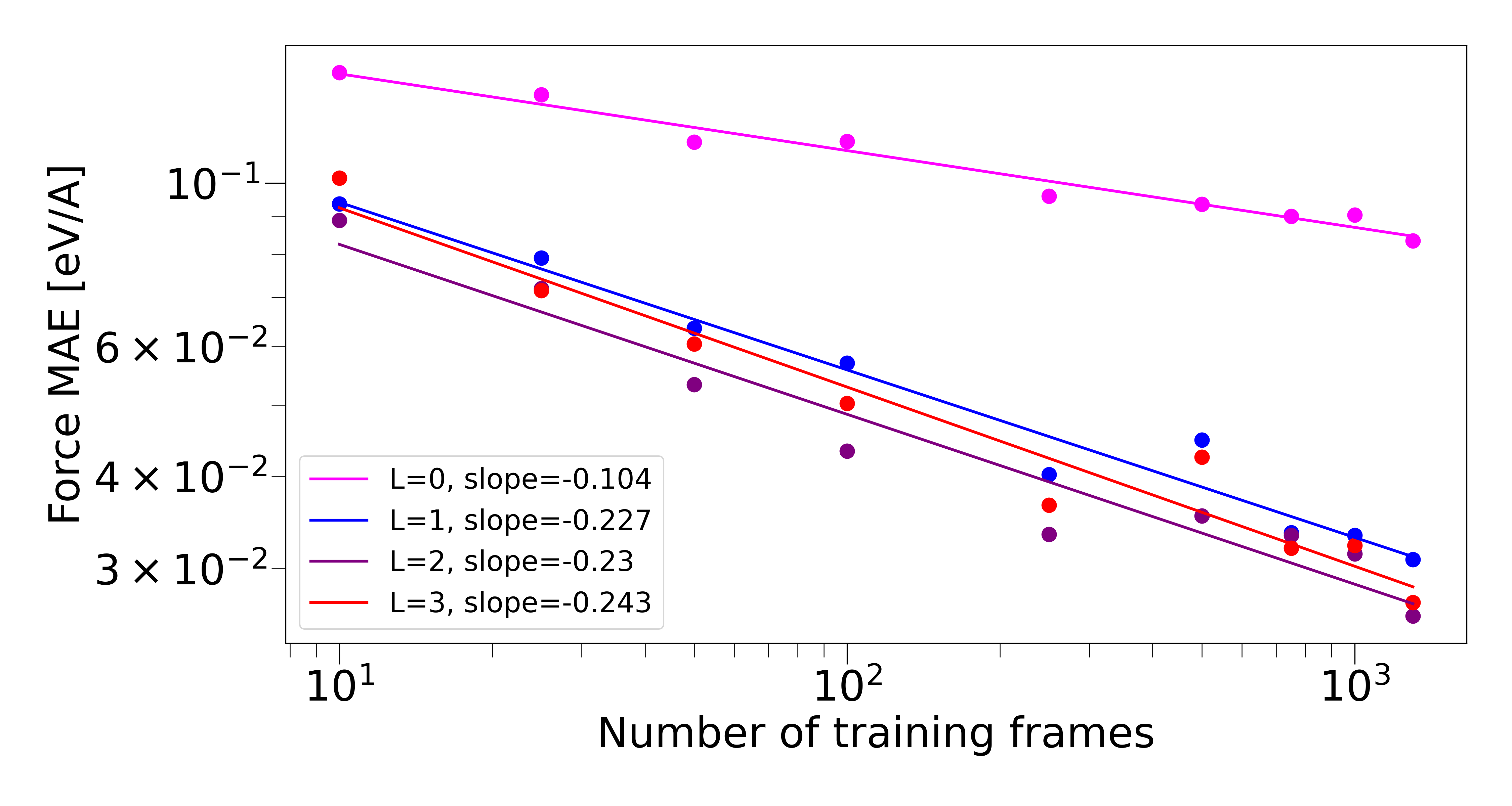}
  \caption{Log-log plot of the predictive error on the water data set from \cite{cheng2019ab} using NequIP with $l \in \{0, 1, 2, 3\}$ as a function of training set size, measured via the force MAE. The equivariant networks display a different scaling behavior than the invariant network.}
  \label{fig:water_data_eff}
\end{figure*}

\begin{figure*}[!htbp]
  \includegraphics[width=\linewidth]{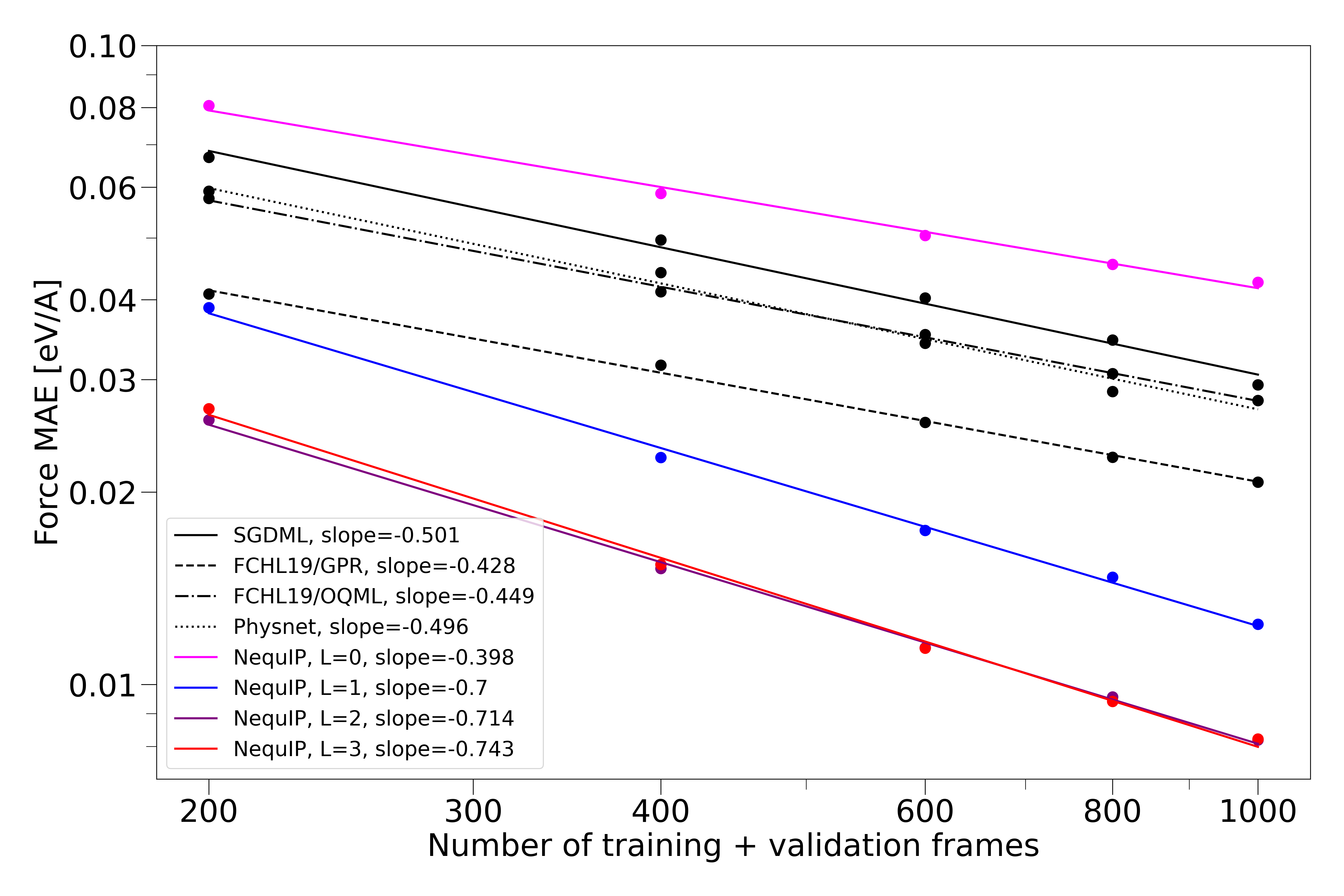}
  \caption{Log-log plot of the predictive error on the aspirin molecule in MD-17 using NequIP with $l \in \{0, 1, 2, 3\}$ as a function of data set size used for training and validation, measured via the force MAE. The plot also shows errors of a series of other methods, including both kernel and deep-learning approaches.}
  \label{fig:md17_data_eff}
\end{figure*}

As a first test system we choose bulk water: in particular we use the data set introduced in \cite{cheng2019ab}. We train a series of networks with identical hyperparameters, but vary the training set sizes between 10 and 1,000 structures. As shown in Figure \ref{fig:water_data_eff}, we find that the equivariant networks with $l \in {1, 2, 3}$ significantly outperform the invariant networks with $l=0$ for all data set sizes as measured by the MAE of force components. This suggests that it is indeed the use of tensor features and equivariant convolutions that enables the high sample efficiency of NequIP. In addition, it is apparent that the learning curves of equivariant networks have a different slope in log-log space. It has been observed that learning curves typically follow a power-law of the form \cite{hestness2017deep}: $\epsilon \propto aN^b$ where $\epsilon$ and $N$ refer to the generalization error and the number of training points, respectively. The exponent of this power-law (or equivalently the slope in log-log space) determines how fast a learning algorithm learns as new data become available. Empirical results have shown that this exponent typically remains fixed across different learning algorithms for a given data set, and different methods only \emph{shift} the learning curve, leaving the log-log slope unaffected \cite{hestness2017deep}. The same trend can also be observed for various methods on the aspirin molecule in the MD-17 data set (see Figure \ref{fig:md17_data_eff}) where across a series of descriptors and regression models (sGDML, FCHL19, and PhysNet \cite{sgdml, christensen2020fchl, physnet_jctc}) the learning curves show an approximately similar log-log slope (results obtained from \url{http://quantum-machine.org/gdml/#datasets}). To our surprise, we observe that the equivariant NequIP networks break this pattern. Instead they follow a log-log slope with larger magnitude, meaning that they learn faster as new data become available. An invariant $l=0$ NequIP network, however, displays a similar log-log slope to other methods, suggesting that it is indeed the equivariant nature of NequIP that allows for the change in learning behavior. Further increasing the rotation order $l$ beyond $l=1$ again only shifts the learning curve and does not results in an additional change in log-log slope. To control for the different number of weights and features in orders of different rotation order $l$, we report weight- and feature-controlled data in Appendix B. Both show qualitatively the same effect. The Appendix also contains results on the behavior of the energies, when trained jointly with forces. For details on the training setup and the control experiments, see the Methods section.\\

We further note, that in \cite{cheng2019ab}, a Behler-Parrinello Neural Network (BPNN) was trained on 1303 structures, yielding a RMSE of $\approx 120\ \operatorname{meV/\AA}$ in forces when evaluated on the remaining 290 structures. We find that NequIP $l=2$ models trained with as little as 100 and 250 data points obtain RMSEs of 129.8 meV/\AA \ and 103.4 meV/\AA \, respectively (note that Figure \ref{fig:water_data_eff} shows the MAE).  This provides further evidence that NequIP exhibits significantly improved data efficiency in comparison with existing methods. 

\section{Discussion}

We demonstrate that the Neural Equivariant Interatomic Potential (NequIP), a new type of graph neural network built on E(3)-equivariant convolutions, exhibits state-of-the-art accuracy and exceptional data efficiency on data sets of small molecules and periodic materials. We isolate that the improvements are due to the introduction of equivariant representations in place of more widely used invariant representations. This raises questions about the optimal way to include symmetry in Machine Learning for molecules and materials. A better understanding of why equivariance enables improved sample efficiency and accuracy is likely to be a fruitful direction towards designing better ML algorithms for the construction of Potential Energy Surfaces. In addition to open questions around the effect of equivariance on accuracy and learning dynamics, a clear theoretical understanding of how the many-body character of interactions arises in Message Passing Interatomic Potentials remains elusive. We expect the proposed method will enable researchers in computational chemistry, physics, biology, and materials science to conduct molecular dynamics simulations of complex reactions and phase transformations at increased accuracy and efficiency.

\section{Methods}

\textbf{Software.} All experiments were run with the \texttt{nequip} software available at \url{github.com/mir-group/nequip} in version 0.3.3, git commit \texttt{50ddbfc31bd44e267b7bb7d2d36d76417b0885ec}. In addition, the \texttt{e3nn} library \cite{mario_geiger_2021_4735637} was used under version 0.3.5, PyTorch under version \texttt{1.9.0} \cite{paszke2019pytorch}, PyTorch Geometric under version 1.7.2 \cite{fey2019fast}, and Python under version 3.9.6.\\

\textbf{Reference Data Sets.} \\

\textit{original MD-17}: MD-17 \cite{chmiela2017machine, schutt2017quantum, sgdml} is a data set of eight small organic molecules, obtained from MD simulations at T=500K and computed at the PBE+vdW-TS level of electronic structure theory, resulting in data set sizes between 133,770 and 993,237 structures. The data set was obtained from \url{http://quantum-machine.org/gdml/#datasets}. For each molecule, we use 950 configurations for training and 50 for validation, sampled uniformly from the full data set, and evaluate the test error on all remaining configurations in the data set. \\

\textit{revised MD-17}: The revised MD-17 data set is a recomputed version of MD-17 obtained at the PBE/def2-SVP level of theory. Using a very tight SCF convergence as well as a very dense DFT integration grid, 100,000 structures \cite{christensen2020role} of the original MD-17 data set were recomputed. The data set can be downloaded at \url{https://figshare.com/articles/dataset/Revised_MD17_dataset_rMD17_/12672038}. For each molecule, we use 950 configurations for training and 50 for validation, sampled uniformly from the full data set, and evaluate the test error on all remaining configurations in the data set.\\

\textit{Molecules@CCSD/CCSD(T)}: The data set of small molecules at CCSD and CCSD(T) accuracy \cite{sgdml} contains positions, energies, and forces for five different small molecules: Asprin (CCSD), Benzene, Malondaldehyde, Toluene, Ethanol (all CCSD(T)). Each data set consists of 1,500 structures with the exception of Ethanol, for which 2,000 structures are available. For more detailed information, we direct the reader to \cite{sgdml}. The data set was obtained from \url{http://quantum-machine.org/gdml/#datasets}. The training/validation set consists of a total of 1,000 molecular structures which we split into 950 for training and 50 for validation (sampled uniformly), and we test the accuracy on all remaining structures (we use the train/test split provided with the data set, but further split the training set into training and validation sets).  \\

\textit{Liquid Water and Ice}: The data set of liquid waters and ice structures \cite{deepmd, deepmd_water_paper} was generated from classical AIMD and path-integral AIMD simulations at different temperatures and pressures, computed with a PBE0-TS functional \cite{deepmd}. The data set contains a total of 140,000 structures, of which 100,000 are liquid water and 20,000 are Ice Ih b),10,000 are Ice Ih c), and another 10,000 are Ice Ih d). The liquid water system consists of 64 $\operatorname{H_2O}$ molecules (192 atoms), while the ice structures consist of 96 $\operatorname{H_2O}$ molecules (288 atoms). We use a validation set of 50 frames and report the test accuracy on all remaining structures in the data set. \\

\textit{Formate decomposition on Cu}: The decomposition process of formate on Cu involves configurations corresponding to the cleavage of the C-H bond, initial and intermediate states (monodentate, bidentate formate on Cu $<110>$) and final states (H ad-atom with a desorbed CO$_2$ in the gas phase). Nudged elastic band (NEB) method was first used to generate an initial reaction path of the C-H bond breaking. 12 short ab initio molecular dynamics, starting from different NEB images, were run to collect a total of 6855 DFT structures. The CP2K \cite{Hutter_Iannuzzi_Schiffmann_VandeVondele_2014} code was employed for the AIMD simulations. Each trajectory was generated with a time step of 0.5 fs and 500 total steps. We train NequIP on 2,500 reference structures sampled uniformly from the full data set of 6,855 structures, use a validation set of 250 structures and evaluate the mean absolute error on all remaining structures. Due to the unbalanced nature of the data set (more atoms of Cu than in the molecule), we use a per-element weighed loss function in which atoms C, $\operatorname{O_1}$, $\operatorname{O_2}$, and H and the sum of all Cu atoms all receive equal weights. We weight the force term with $N_{atoms}^2 = 2,704$ and the energy term with 1. \\

\textit{Li\textsubscript{4}P\textsubscript{2}O\textsubscript{7} glass}: The Li\textsubscript{4}P\textsubscript{2}O\textsubscript{7} ab-initio data were generated using an ab-initio melt-quench MD simulation, starting with a stoichiometric crystal of 208 atoms (space group P21/c) in a periodic box of $10.4 \times 14.0 \times 16.0$ \AA.
The dynamics used the Vienna Ab-Initio Simulation Package (VASP) \cite{vasp1,vasp2,vasp3},
with a generalized gradient PBE functional \cite{PBE}, projector augmented wave (PAW)
pseudopotentials \cite{PAW},
a NVT ensemble and a Nos\'e-Hoover thermostat,
a time step of 2 fs,
a plane-wave cutoff of 400 eV,
and a $\Gamma$-point reciprocal-space mesh.
The crystal was melted at 3000 K for 50 ps, then immediately quenched to 600 K and run for another 50 ps. The resulting structure was confirmed to be amorphous by plotting the radial distribution function of P-P distances.
The training was performed only on the molten portion, and the MD simulations for a quenched simulation. We sample the training sets uniformly from the full data set of 25,000 AIMD frames. We use a validation set of 100 structures, and evaluate the model on all remaining structures of the melt trajectory as well as on the full quench trajectory.\\

\textit{LiPS:} Lithium phosphorus sulfide (LiPS) based materials are known to exhibit high lithium ion conductivity, making them attractive as solid-state electrolytes for lithium-ion batteries. Other examples of known materials in this family of superionic conductors are LiGePS and LiCuPS-based compounds. The training data set is taken from a previous study on graph neural network force field \cite{gnnff}, where the LiPS training data were generated using ab-initio MD of an LiPS structure with Li-vacancy (Li\textsubscript{6.75}P\textsubscript{3}S\textsubscript{11}) consisting of 27 Li, 12 P, and 44 S atoms respectively. The structure was first equilibrated and then run at 520 K using the NVT ensemble for 50 ps with a 2.0 fs time step. The full data set contains 25,001 MD frames. We choose training set sizes of 10, 100, 1,000, and 2,500 frames with a fixed validation set size of 100. \\

\textit{Liquid Water, Cheng et al.}: The training set used in the data efficiency experiments on water consists of 1,593 reference calculations of bulk liquid water at the revPBE0-D3 level of accuracy, with each structure containing 192 atoms, as given in \cite{cheng2019ab}. Further information can be found in \cite{cheng2019ab}. The data set was obtained from \url{https://github.com/BingqingCheng/ab-initio-thermodynamics-of-water}. We sample the training set uniformly from the full data set and for each experiment also use a validation set consisting of 100 structures. We then evaluate the error on a fixed hold-out test set of 190 structures.\\

\textbf{Molecular Dynamics Simulations.} To run MD simulations, NequIP force outputs were integrated with the Atomic Simulation Environment (ASE) \cite{Hjorth_Larsen_2017} in which we implement a custom version of the Nosé-Hoover thermostat. We use this in-house implementation for the both the Li\textsubscript{4}P\textsubscript{2}O\textsubscript{7} as well as the LiPS MD simulations. The thermostat parameter was chosen to match the temperature fluctuations observed in the AIMD run. The RDF and ADFs for Li\textsubscript{4}P\textsubscript{2}O\textsubscript{7} were computed with a maximum distance of 6 \AA \ (RDF) and 2.5 \AA \ (both ADFs). The Li\textsubscript{4}P\textsubscript{2}O\textsubscript{7} MD simulations were started from the first frame of the AIMD quench simulation and the LiPS simulation was started from the first frame of the reference AIMD simulation of the corresponding training data.\\

\textbf{Training.} Networks are trained using a loss function based on a weighted sum of energy and a force loss terms:

\begin{equation}
    \mathcal{L} = \lambda_E || \hat{E} - E||^2 + \lambda_F \frac{1}{3N} \sum_{i=1}^{N} \sum_{\alpha=1}^3 || -\frac{\partial \hat{E}}{\partial r_{i, \alpha}}  - F_{i, \alpha}||^2
\end{equation}

where N is the number of atoms in the system, $\hat{E}$ is the predicted potential energy, and $\lambda_E$ and $\lambda_F$ are the energy- and force-weightings, respectively. While it is helpful to optimize the weightings as a hyperparameter, we found a relative weighting of energies to forces of $1$ to $N_{atoms}^2$ a suitable default choice. Here the $N$ accounts for the fact that that potential energy is a \emph{global} quantity, while the atomic forces are \emph{local} quantities and the square accounts for the fact that we use a MSE loss. This also makes the loss function size invariant. A full set of the weightings used in this work can be found in table \ref{tab:hyper}. \\ 

We normalize the target energies by subtracting the mean potential energy over the training set and scale both the target energies and target force components by the root mean square of the force components over the training set. The predicted atomic energies $\hat{E}_i$ are scaled and shifted by two learnable per-species parameters before summing them for the total predicted potential energy $\hat{E}$: 

\begin{equation}
    \hat{E} = \sum_{i}{\sigma_{s_i}\hat{E}_{i} + \lambda_{s_i}}
\end{equation}

where $\sigma_{s_i}$ and $\lambda_{s_i}$ are learnable per-species parameters indexed by $s_i$, the species of atom $i$. They are initialized to 1 and 0, respectively. 

For the case of the joint training on water and ice, since the liquid water and ice structures have different numbers of atoms, we do not scale or shift the potential energy targets or force targets. Instead, we initialize the learnable per-species shift to the mean per-atom energy and initialize the learnable per-species scale to the average standard deviation over all force components in the training set.\\ 

\textbf{Learning Curve Experiments} For learning curve experiments on the aspirin molecule in MD-17, a series of NequIP models with increasing order $l \in \{0, 1, 2, 3\}$ were trained on varying data set sizes. In particular, experiments were performed with a budget for training and validation of 200, 400, 600, 800, 1000 configurations, of which 50 samples were used for validation while the remaining ones were used for training. The reported test error was computed on the entire remaining MD-17 trajectory for each given budget. The weight-controlled version of NequIP was set up by creating a $l=0$ network with increased feature size that matches the number of weights up to approx. $0.1\%$ of the $l=1$ network. The feature-controlled version of NequIP was set up by creating a $l=0$ network with the same number of features as the $l=1$ network, i.e. 4x more features than the original $l=0$ network (1 scalar and 3 vector features), in particular the $l=1$ network had a feature configuration of \texttt{64x0o + 64x0e + 64x1o + 64x1e} while the original $l=0$ network used \texttt{64x0e} and feature-controlled $l=0$ network used \texttt{512x0e}.\\

\textbf{Hyperparameters.} All models were trained on a NVIDIA Tesla V100 GPU in single-GPU training. For the small molecule systems, we use 5 interaction blocks, a learning rate of 0.01 and a batch size of 5. For the periodic systems, we use 6 interaction blocks, a learning rate of 0.005 and a batch size of 1.  We decrease the initial learning rate by a decay factor of 0.8 whenever the validation loss in the forces has not seen an improvement for 50 epochs. We continuously save the model with the best validation loss in the forces and use the model with the overall best validation loss for evaluation on the test set and MD simulations. For validation and test error evaluation, we use an exponential moving average of the training weights with weight 0.99. Training is stopped if either of the following conditions is met: a) a maximum training time of of 7 days is reached; b) a maximum number 1,000,000 epochs is reached; c) the learning rate drops below $10^{-6}$; d) the validation loss does not improve for 1,000 epochs. We note that competitive results can typically be obtained within a matter of hours or often even minutes and most of the remaining training time is spent on only small improvements in the errors. We found the use of small batch sizes to be an important hyperparameter. We also found it important to choose the radial cutoff distance $r_c$ appropriately for a given system. In addition, we observed the number of layers to not have a strong effect as long as they were set within a reasonable range. We use different numbers of $l$ and feature dimensions for different systems and similarly also vary the cutoff radius for different systems. A full outline of the choices for $l$, feature size, cutoff radius as well as the weights for energies and forces in the loss function can be found in \ref{tab:hyper}. All models were trained with both even and odd features. The weights were initialized according to a standard normal distribution (for details, see the \texttt{e3nn} software implementation \cite{mario_geiger_2021_4735637}). The invariant radial networks act on a trainable Bessel basis of size 8 and were implemented with 3 hidden layers of 64 neurons with SiLU nonlinearities between them. The even scalars of the final interaction block are passed to the output block, which  first reduces the feature dimension to 16 even scalars through a self-interaction layer. Finally, through another self-interaction layer, the feature dimension is reduced to a single scalar output value associated with each atom which is then summed over to give the total potential energy. Forces are obtained as the negative gradient of this predicted total potential energy, computed via automatic differentiation. All models were optimized with Adam with the AMSGrad variant in the PyTorch implementation \cite{kingma2014adam, loshchilov2017decoupled, reddi2019convergence} with $\beta_1=0.9$, $\beta_2=0.999$, and $\epsilon=10^{-8}$ without weight decay. The average number of neighbors used for the $\frac{1}{\sqrt{N}}$ normalization of the convolution was computed over the full training set. For all molecular results, the average number of neighbors was computed once on the N=1000 case for revised MD-17 and used for all other experiments. For the water sample efficiency and the LiPS experiments it was computed once on the N=1000  and N=2500 cases, respectively and then used for all other experiments for that system. All input files for training of NequIP models will be shared upon publication. 

\begin{table*}[!htbp]
\centering
\begin{tabular}{lccccc}
\hline \hline
Data Set  & Tensor rank $l$ & \# Features &  $r_c$ & $\lambda_E$ & $\lambda_F$ \\
\hline
MD-17  & 3 & 64 & 4.0 &  1  & 1,000 \\
revMD-17   & \{0, 1, 2, 3\} & 64 & 4.0 &  1 & 1,000 \\
CCSD/CCSD(T)  & 3 &  64 & 4.0 &   1 & 1,000\\
Water+Ices, DeepMD  & 2 & 32  & 6.0 & see \ref{tab:deepmd} & see \ref{tab:deepmd}  \\ 
Formate on Cu  & 2 & 32 & 5.0 &  1 & 2,704 \\
Li\textsubscript{4}P\textsubscript{2}O\textsubscript{7}  & 2  & 32  & 5.0. & 1  & 43,264 \\
LiPS  &  2&32  & 5.0 & 1  & 6,889 \\
Water, Cheng et al. & \{0, 1, 2, 3\} & 32  & 4.5 & 1 &  36,864 \\
\hline \hline
\end{tabular}
\caption{Tensor rank $l$, feature size, radial cutoff in units of [\AA], as well as energy and force weights used in the joint loss function. All models were trained with even and odd features, i.e. a tensor rank of $l=1$ and 32 features corresponds to $32x0\mathrm{o} + 32x0\mathrm{e} + 32x1\mathrm{o} + 32x1\mathrm{e}$. The force weightings for formate on Cu, LiPO, LiPS, and the water system for sample efficiency tests stem from $N_{atoms}^2$. 
\label{tab:hyper}}
\end{table*}

\section{Data Availability}

An open-source software implementation of NequIP is available at \url{https://github.com/mir-group/nequip}. All data sets will be made available upon publication. 
\bibliographystyle{naturemag}
\bibliography{bib.bib}

\begin{thebibliography}{10}
\expandafter\ifx\csname url\endcsname\relax
  \def\url#1{\texttt{#1}}\fi
\expandafter\ifx\csname urlprefix\endcsname\relax\def\urlprefix{URL }\fi
\providecommand{\bibinfo}[2]{#2}
\providecommand{\eprint}[2][]{\url{#2}}

\bibitem{richards2016design}
\bibinfo{author}{Richards, W.~D.} \emph{et~al.}
\newblock \bibinfo{title}{Design and synthesis of the superionic conductor na
  10 snp 2 s 12}.
\newblock \emph{\bibinfo{journal}{Nature communications}}
  \textbf{\bibinfo{volume}{7}}, \bibinfo{pages}{1--8} (\bibinfo{year}{2016}).

\bibitem{boero1998first}
\bibinfo{author}{Boero, M.}, \bibinfo{author}{Parrinello, M.} \&
  \bibinfo{author}{Terakura, K.}
\newblock \bibinfo{title}{First principles molecular dynamics study of ziegler-
  natta heterogeneous catalysis}.
\newblock \emph{\bibinfo{journal}{Journal of the American Chemical Society}}
  \textbf{\bibinfo{volume}{120}}, \bibinfo{pages}{2746--2752}
  (\bibinfo{year}{1998}).

\bibitem{lindorff2011fast}
\bibinfo{author}{Lindorff-Larsen, K.}, \bibinfo{author}{Piana, S.},
  \bibinfo{author}{Dror, R.~O.} \& \bibinfo{author}{Shaw, D.~E.}
\newblock \bibinfo{title}{How fast-folding proteins fold}.
\newblock \emph{\bibinfo{journal}{Science}} \textbf{\bibinfo{volume}{334}},
  \bibinfo{pages}{517--520} (\bibinfo{year}{2011}).

\bibitem{behler2007generalized}
\bibinfo{author}{Behler, J.} \& \bibinfo{author}{Parrinello, M.}
\newblock \bibinfo{title}{Generalized neural-network representation of
  high-dimensional potential-energy surfaces}.
\newblock \emph{\bibinfo{journal}{Physical review letters}}
  \textbf{\bibinfo{volume}{98}}, \bibinfo{pages}{146401}
  (\bibinfo{year}{2007}).

\bibitem{gaporiginalpaper}
\bibinfo{author}{Bart{\'o}k, A.~P.}, \bibinfo{author}{Payne, M.~C.},
  \bibinfo{author}{Kondor, R.} \& \bibinfo{author}{Cs{\'a}nyi, G.}
\newblock \bibinfo{title}{Gaussian approximation potentials: The accuracy of
  quantum mechanics, without the electrons}.
\newblock \emph{\bibinfo{journal}{Physical review letters}}
  \textbf{\bibinfo{volume}{104}}, \bibinfo{pages}{136403}
  (\bibinfo{year}{2010}).

\bibitem{shapeev2016moment}
\bibinfo{author}{Shapeev, A.~V.}
\newblock \bibinfo{title}{Moment tensor potentials: A class of systematically
  improvable interatomic potentials}.
\newblock \emph{\bibinfo{journal}{Multiscale Modeling \& Simulation}}
  \textbf{\bibinfo{volume}{14}}, \bibinfo{pages}{1153--1173}
  (\bibinfo{year}{2016}).

\bibitem{thompson2015spectral}
\bibinfo{author}{Thompson, A.~P.}, \bibinfo{author}{Swiler, L.~P.},
  \bibinfo{author}{Trott, C.~R.}, \bibinfo{author}{Foiles, S.~M.} \&
  \bibinfo{author}{Tucker, G.~J.}
\newblock \bibinfo{title}{Spectral neighbor analysis method for automated
  generation of quantum-accurate interatomic potentials}.
\newblock \emph{\bibinfo{journal}{Journal of Computational Physics}}
  \textbf{\bibinfo{volume}{285}}, \bibinfo{pages}{316--330}
  (\bibinfo{year}{2015}).

\bibitem{vandermause2020fly}
\bibinfo{author}{Vandermause, J.} \emph{et~al.}
\newblock \bibinfo{title}{On-the-fly active learning of interpretable bayesian
  force fields for atomistic rare events}.
\newblock \emph{\bibinfo{journal}{npj Computational Materials}}
  \textbf{\bibinfo{volume}{6}}, \bibinfo{pages}{1--11} (\bibinfo{year}{2020}).

\bibitem{schnet_neurips}
\bibinfo{author}{Sch{\"u}tt, K.} \emph{et~al.}
\newblock \bibinfo{title}{Schnet: A continuous-filter convolutional neural
  network for modeling quantum interactions}.
\newblock In \emph{\bibinfo{booktitle}{Advances in neural information
  processing systems}}, \bibinfo{pages}{991--1001} (\bibinfo{year}{2017}).

\bibitem{physnet_jctc}
\bibinfo{author}{Unke, O.~T.} \& \bibinfo{author}{Meuwly, M.}
\newblock \bibinfo{title}{Physnet: A neural network for predicting energies,
  forces, dipole moments, and partial charges}.
\newblock \emph{\bibinfo{journal}{Journal of chemical theory and computation}}
  \textbf{\bibinfo{volume}{15}}, \bibinfo{pages}{3678--3693}
  (\bibinfo{year}{2019}).

\bibitem{klicpera2020directional}
\bibinfo{author}{Klicpera, J.}, \bibinfo{author}{Gro{\ss}, J.} \&
  \bibinfo{author}{G{\"u}nnemann, S.}
\newblock \bibinfo{title}{Directional message passing for molecular graphs}.
\newblock \emph{\bibinfo{journal}{arXiv preprint arXiv:2003.03123}}
  (\bibinfo{year}{2020}).

\bibitem{dcf}
\bibinfo{author}{Mailoa, J.~P.} \emph{et~al.}
\newblock \bibinfo{title}{A fast neural network approach for direct covariant
  forces prediction in complex multi-element extended systems}.
\newblock \emph{\bibinfo{journal}{Nature machine intelligence}}
  \textbf{\bibinfo{volume}{1}}, \bibinfo{pages}{471--479}
  (\bibinfo{year}{2019}).

\bibitem{gnnff}
\bibinfo{author}{Park, C.~W.} \emph{et~al.}
\newblock \bibinfo{title}{Accurate and scalable multi-element graph neural
  network force field and molecular dynamics with direct force architecture}.
\newblock \emph{\bibinfo{journal}{arXiv preprint arXiv:2007.14444}}
  (\bibinfo{year}{2020}).

\bibitem{artrith2014understanding}
\bibinfo{author}{Artrith, N.} \& \bibinfo{author}{Kolpak, A.~M.}
\newblock \bibinfo{title}{Understanding the composition and activity of
  electrocatalytic nanoalloys in aqueous solvents: A combination of dft and
  accurate neural network potentials}.
\newblock \emph{\bibinfo{journal}{Nano letters}} \textbf{\bibinfo{volume}{14}},
  \bibinfo{pages}{2670--2676} (\bibinfo{year}{2014}).

\bibitem{deepmd}
\bibinfo{author}{Zhang, L.}, \bibinfo{author}{Han, J.}, \bibinfo{author}{Wang,
  H.}, \bibinfo{author}{Car, R.} \& \bibinfo{author}{Weinan, E.}
\newblock \bibinfo{title}{Deep potential molecular dynamics: a scalable model
  with the accuracy of quantum mechanics}.
\newblock \emph{\bibinfo{journal}{Physical review letters}}
  \textbf{\bibinfo{volume}{120}}, \bibinfo{pages}{143001}
  (\bibinfo{year}{2018}).

\bibitem{smith2017ani}
\bibinfo{author}{Smith, J.~S.}, \bibinfo{author}{Isayev, O.} \&
  \bibinfo{author}{Roitberg, A.~E.}
\newblock \bibinfo{title}{Ani-1: an extensible neural network potential with
  dft accuracy at force field computational cost}.
\newblock \emph{\bibinfo{journal}{Chemical science}}
  \textbf{\bibinfo{volume}{8}}, \bibinfo{pages}{3192--3203}
  (\bibinfo{year}{2017}).

\bibitem{gilmer2017neural}
\bibinfo{author}{Gilmer, J.}, \bibinfo{author}{Schoenholz, S.~S.},
  \bibinfo{author}{Riley, P.~F.}, \bibinfo{author}{Vinyals, O.} \&
  \bibinfo{author}{Dahl, G.~E.}
\newblock \bibinfo{title}{Neural message passing for quantum chemistry}.
\newblock \emph{\bibinfo{journal}{arXiv preprint arXiv:1704.01212}}
  (\bibinfo{year}{2017}).

\bibitem{anderson2019cormorant}
\bibinfo{author}{Anderson, B.}, \bibinfo{author}{Hy, T.~S.} \&
  \bibinfo{author}{Kondor, R.}
\newblock \bibinfo{title}{Cormorant: Covariant molecular neural networks}.
\newblock In \emph{\bibinfo{booktitle}{Advances in Neural Information
  Processing Systems}}, \bibinfo{pages}{14537--14546} (\bibinfo{year}{2019}).

\bibitem{townshend2020geometric}
\bibinfo{author}{Townshend, R.~J.}, \bibinfo{author}{Townshend, B.},
  \bibinfo{author}{Eismann, S.} \& \bibinfo{author}{Dror, R.~O.}
\newblock \bibinfo{title}{Geometric prediction: Moving beyond scalars}.
\newblock \emph{\bibinfo{journal}{arXiv preprint arXiv:2006.14163}}
  (\bibinfo{year}{2020}).

\bibitem{thomas2018tensor}
\bibinfo{author}{Thomas, N.} \emph{et~al.}
\newblock \bibinfo{title}{Tensor field networks: Rotation-and
  translation-equivariant neural networks for 3d point clouds}.
\newblock \emph{\bibinfo{journal}{arXiv preprint arXiv:1802.08219}}
  (\bibinfo{year}{2018}).

\bibitem{nequip-arxiv-v1}
\bibinfo{author}{Batzner, S.} \emph{et~al.}
\newblock \bibinfo{title}{Se(3)-equivariant graph neural networks for
  data-efficient and accurate interatomic potentials}.
\newblock \emph{\bibinfo{journal}{arXiv preprint arXiv:2101.03164v1}}
  (\bibinfo{year}{2021}).

\bibitem{schutt2021equivariant}
\bibinfo{author}{Sch{\"u}tt, K.~T.}, \bibinfo{author}{Unke, O.~T.} \&
  \bibinfo{author}{Gastegger, M.}
\newblock \bibinfo{title}{Equivariant message passing for the prediction of
  tensorial properties and molecular spectra}.
\newblock \emph{\bibinfo{journal}{arXiv preprint arXiv:2102.03150}}
  (\bibinfo{year}{2021}).

\bibitem{haghighatlari2021newtonnet}
\bibinfo{author}{Haghighatlari, M.} \emph{et~al.}
\newblock \bibinfo{title}{Newtonnet: A newtonian message passing network for
  deep learning of interatomic potentials and forces}.
\newblock \emph{\bibinfo{journal}{arXiv preprint arXiv:2108.02913}}
  (\bibinfo{year}{2021}).

\bibitem{klicpera2021gemnet}
\bibinfo{author}{Klicpera, J.}, \bibinfo{author}{Becker, F.} \&
  \bibinfo{author}{G{\"u}nnemann, S.}
\newblock \bibinfo{title}{Gemnet: Universal directional graph neural networks
  for molecules}.
\newblock \emph{\bibinfo{journal}{arXiv preprint arXiv:2106.08903}}
  (\bibinfo{year}{2021}).

\bibitem{unke2021}
\bibinfo{author}{Unke, O.~T.} \emph{et~al.}
\newblock \bibinfo{title}{Spookynet: Learning force fields with electronic
  degrees of freedom and nonlocal effects}.
\newblock \emph{\bibinfo{journal}{Nature Communications}}
  \textbf{\bibinfo{volume}{12}}, \bibinfo{pages}{7273} (\bibinfo{year}{2021}).
\newblock \urlprefix\url{https://doi.org/10.1038/s41467-021-27504-0}.

\bibitem{qiao2021unite}
\bibinfo{author}{Qiao, Z.} \emph{et~al.}
\newblock \bibinfo{title}{Unite: Unitary n-body tensor equivariant network with
  applications to quantum chemistry}.
\newblock \emph{\bibinfo{journal}{arXiv preprint arXiv:2105.14655}}
  (\bibinfo{year}{2021}).

\bibitem{grisafi2019atomic}
\bibinfo{author}{Grisafi, A.}, \bibinfo{author}{Wilkins, D.~M.},
  \bibinfo{author}{Willatt, M.~J.} \& \bibinfo{author}{Ceriotti, M.}
\newblock \bibinfo{title}{Atomic-scale representation and statistical learning
  of tensorial properties}.
\newblock In \emph{\bibinfo{booktitle}{Machine Learning in Chemistry:
  Data-Driven Algorithms, Learning Systems, and Predictions}},
  \bibinfo{pages}{1--21} (\bibinfo{publisher}{ACS Publications},
  \bibinfo{year}{2019}).

\bibitem{weiler20183d}
\bibinfo{author}{Weiler, M.}, \bibinfo{author}{Geiger, M.},
  \bibinfo{author}{Welling, M.}, \bibinfo{author}{Boomsma, W.} \&
  \bibinfo{author}{Cohen, T.~S.}
\newblock \bibinfo{title}{3d steerable cnns: Learning rotationally equivariant
  features in volumetric data}.
\newblock In \emph{\bibinfo{booktitle}{Advances in Neural Information
  Processing Systems}}, \bibinfo{pages}{10381--10392} (\bibinfo{year}{2018}).

\bibitem{kondor2018n}
\bibinfo{author}{Kondor, R.}
\newblock \bibinfo{title}{N-body networks: a covariant hierarchical neural
  network architecture for learning atomic potentials}.
\newblock \emph{\bibinfo{journal}{arXiv preprint arXiv:1803.01588}}
  (\bibinfo{year}{2018}).

\bibitem{kondor2018clebsch}
\bibinfo{author}{Kondor, R.}, \bibinfo{author}{Lin, Z.} \&
  \bibinfo{author}{Trivedi, S.}
\newblock \bibinfo{title}{Clebsch--gordan nets: a fully fourier space spherical
  convolutional neural network}.
\newblock In \emph{\bibinfo{booktitle}{Advances in Neural Information
  Processing Systems}}, \bibinfo{pages}{10117--10126} (\bibinfo{year}{2018}).

\bibitem{mario_geiger_2021_4735637}
\bibinfo{author}{Geiger, M.} \emph{et~al.}
\newblock \bibinfo{title}{e3nn/e3nn: 2021-05-04} (\bibinfo{year}{2021}).
\newblock \urlprefix\url{https://doi.org/10.5281/zenodo.4735637}.

\bibitem{hendrycks2016gaussian}
\bibinfo{author}{Hendrycks, D.} \& \bibinfo{author}{Gimpel, K.}
\newblock \bibinfo{title}{Gaussian error linear units (gelus)}.
\newblock \emph{\bibinfo{journal}{arXiv preprint arXiv:1606.08415}}
  (\bibinfo{year}{2016}).

\bibitem{Hjorth_Larsen_2017}
\bibinfo{author}{Larsen, A.~H.} \emph{et~al.}
\newblock \bibinfo{title}{The atomic simulation environment{\textemdash}a
  python library for working with atoms}.
\newblock \emph{\bibinfo{journal}{Journal of Physics: Condensed Matter}}
  \textbf{\bibinfo{volume}{29}}, \bibinfo{pages}{273002}
  (\bibinfo{year}{2017}).
\newblock \urlprefix\url{https://doi.org/10.1088%2F1361-648x%2Faa680e}.

\bibitem{he2016deep}
\bibinfo{author}{He, K.}, \bibinfo{author}{Zhang, X.}, \bibinfo{author}{Ren,
  S.} \& \bibinfo{author}{Sun, J.}
\newblock \bibinfo{title}{Deep residual learning for image recognition}.
\newblock In \emph{\bibinfo{booktitle}{Proceedings of the IEEE conference on
  computer vision and pattern recognition}}, \bibinfo{pages}{770--778}
  (\bibinfo{year}{2016}).

\bibitem{chmiela2017machine}
\bibinfo{author}{Chmiela, S.} \emph{et~al.}
\newblock \bibinfo{title}{Machine learning of accurate energy-conserving
  molecular force fields}.
\newblock \emph{\bibinfo{journal}{Science advances}}
  \textbf{\bibinfo{volume}{3}}, \bibinfo{pages}{e1603015}
  (\bibinfo{year}{2017}).

\bibitem{schutt2017quantum}
\bibinfo{author}{Sch{\"u}tt, K.~T.}, \bibinfo{author}{Arbabzadah, F.},
  \bibinfo{author}{Chmiela, S.}, \bibinfo{author}{M{\"u}ller, K.~R.} \&
  \bibinfo{author}{Tkatchenko, A.}
\newblock \bibinfo{title}{Quantum-chemical insights from deep tensor neural
  networks}.
\newblock \emph{\bibinfo{journal}{Nature Communications}}
  \textbf{\bibinfo{volume}{8}}, \bibinfo{pages}{13890} (\bibinfo{year}{2017}).
\newblock \urlprefix\url{https://doi.org/10.1038/ncomms13890}.

\bibitem{sgdml}
\bibinfo{author}{Chmiela, S.}, \bibinfo{author}{Sauceda, H.~E.},
  \bibinfo{author}{Müller, K.-R.} \& \bibinfo{author}{Tkatchenko, A.}
\newblock \bibinfo{title}{Towards exact molecular dynamics simulations with
  machine-learned force fields}.
\newblock \emph{\bibinfo{journal}{Nature Communications}}
  \textbf{\bibinfo{volume}{9}}, \bibinfo{pages}{3887} (\bibinfo{year}{2018}).

\bibitem{deepmd_water_paper}
\bibinfo{author}{Ko, H.-Y.} \emph{et~al.}
\newblock \bibinfo{title}{Isotope effects in liquid water via deep potential
  molecular dynamics}.
\newblock \emph{\bibinfo{journal}{Molecular Physics}}
  \textbf{\bibinfo{volume}{117}}, \bibinfo{pages}{3269--3281}
  (\bibinfo{year}{2019}).

\bibitem{christensen2020role}
\bibinfo{author}{Christensen, A.~S.} \& \bibinfo{author}{von Lilienfeld, O.~A.}
\newblock \bibinfo{title}{On the role of gradients for machine learning of
  molecular energies and forces}.
\newblock \emph{\bibinfo{journal}{Machine Learning: Science and Technology}}
  \textbf{\bibinfo{volume}{1}}, \bibinfo{pages}{045018} (\bibinfo{year}{2020}).

\bibitem{devereux2020extending}
\bibinfo{author}{Devereux, C.} \emph{et~al.}
\newblock \bibinfo{title}{Extending the applicability of the ani deep learning
  molecular potential to sulfur and halogens}.
\newblock \emph{\bibinfo{journal}{Journal of Chemical Theory and Computation}}
  \textbf{\bibinfo{volume}{16}}, \bibinfo{pages}{4192--4202}
  (\bibinfo{year}{2020}).

\bibitem{christensen2020fchl}
\bibinfo{author}{Christensen, A.~S.}, \bibinfo{author}{Bratholm, L.~A.},
  \bibinfo{author}{Faber, F.~A.} \& \bibinfo{author}{Anatole~von Lilienfeld,
  O.}
\newblock \bibinfo{title}{Fchl revisited: Faster and more accurate quantum
  machine learning}.
\newblock \emph{\bibinfo{journal}{The Journal of Chemical Physics}}
  \textbf{\bibinfo{volume}{152}}, \bibinfo{pages}{044107}
  (\bibinfo{year}{2020}).

\bibitem{drautz2019atomic}
\bibinfo{author}{Drautz, R.}
\newblock \bibinfo{title}{Atomic cluster expansion for accurate and
  transferable interatomic potentials}.
\newblock \emph{\bibinfo{journal}{Physical Review B}}
  \textbf{\bibinfo{volume}{99}}, \bibinfo{pages}{014104}
  (\bibinfo{year}{2019}).

\bibitem{kovacs2021linear}
\bibinfo{author}{Kov{\'a}cs, D.~P.} \emph{et~al.}
\newblock \bibinfo{title}{Linear atomic cluster expansion force fields for
  organic molecules: beyond rmse}.
\newblock \emph{\bibinfo{journal}{Journal of Chemical Theory and Computation}}
  (\bibinfo{year}{2021}).

\bibitem{zhang2018end}
\bibinfo{author}{Zhang, L.} \emph{et~al.}
\newblock \bibinfo{title}{End-to-end symmetry preserving inter-atomic potential
  energy model for finite and extended systems}.
\newblock \emph{\bibinfo{journal}{Advances in Neural Information Processing
  Systems}} \textbf{\bibinfo{volume}{31}} (\bibinfo{year}{2018}).

\bibitem{Sim_Gardner_King_1996}
\bibinfo{author}{Sim, W.~S.}, \bibinfo{author}{Gardner, P.} \&
  \bibinfo{author}{King, D.~A.}
\newblock \bibinfo{title}{Multiple bonding configurations of adsorbed formate
  on ag{111}}.
\newblock \emph{\bibinfo{journal}{The Journal of Physical Chemistry}}
  \textbf{\bibinfo{volume}{100}}, \bibinfo{pages}{12509–12516}
  (\bibinfo{year}{1996}).
\newblock \bibinfo{note}{00000}.

\bibitem{Wang_Morikawa_Matsumoto_Nakamura_2006}
\bibinfo{author}{Wang, G.}, \bibinfo{author}{Morikawa, Y.},
  \bibinfo{author}{Matsumoto, T.} \& \bibinfo{author}{Nakamura, J.}
\newblock \bibinfo{title}{Why is formate synthesis insensitive to copper
  surface structures?}
\newblock \emph{\bibinfo{journal}{The Journal of Physical Chemistry B}}
  \textbf{\bibinfo{volume}{110}}, \bibinfo{pages}{9–11}
  (\bibinfo{year}{2006}).
\newblock \bibinfo{note}{00050}.

\bibitem{LiPON}
\bibinfo{author}{Yu, X.}, \bibinfo{author}{Bates, J.~B.},
  \bibinfo{author}{Jellison, G.~E.} \& \bibinfo{author}{Hart, F.~X.}
\newblock \bibinfo{title}{A stable thin-film lithium electrolyte: Lithium
  phosphorus oxynitride}.
\newblock \emph{\bibinfo{journal}{Journal of The Electrochemical Society}}
  \textbf{\bibinfo{volume}{144}}, \bibinfo{pages}{524--532}
  (\bibinfo{year}{1997}).
\newblock \urlprefix\url{https://doi.org/10.1149/1.1837443}.

\bibitem{LiSiPON}
\bibinfo{author}{Westover, A.~S.} \emph{et~al.}
\newblock \bibinfo{title}{Plasma synthesis of spherical crystalline and
  amorphous electrolyte nanopowders for solid-state batteries}.
\newblock \emph{\bibinfo{journal}{{ACS} Applied Materials {\&} Interfaces}}
  \textbf{\bibinfo{volume}{12}}, \bibinfo{pages}{11570--11578}
  (\bibinfo{year}{2020}).
\newblock \urlprefix\url{https://doi.org/10.1021/acsami.9b20812}.

\bibitem{li2017study}
\bibinfo{author}{Li, W.}, \bibinfo{author}{Ando, Y.},
  \bibinfo{author}{Minamitani, E.} \& \bibinfo{author}{Watanabe, S.}
\newblock \bibinfo{title}{Study of li atom diffusion in amorphous li3po4 with
  neural network potential}.
\newblock \emph{\bibinfo{journal}{The Journal of chemical physics}}
  \textbf{\bibinfo{volume}{147}}, \bibinfo{pages}{214106}
  (\bibinfo{year}{2017}).

\bibitem{cheng2019ab}
\bibinfo{author}{Cheng, B.}, \bibinfo{author}{Engel, E.~A.},
  \bibinfo{author}{Behler, J.}, \bibinfo{author}{Dellago, C.} \&
  \bibinfo{author}{Ceriotti, M.}
\newblock \bibinfo{title}{Ab initio thermodynamics of liquid and solid water}.
\newblock \emph{\bibinfo{journal}{Proceedings of the National Academy of
  Sciences}} \textbf{\bibinfo{volume}{116}}, \bibinfo{pages}{1110--1115}
  (\bibinfo{year}{2019}).

\bibitem{hestness2017deep}
\bibinfo{author}{Hestness, J.} \emph{et~al.}
\newblock \bibinfo{title}{Deep learning scaling is predictable, empirically}.
\newblock \emph{\bibinfo{journal}{arXiv preprint arXiv:1712.00409}}
  (\bibinfo{year}{2017}).

\bibitem{paszke2019pytorch}
\bibinfo{author}{Paszke, A.} \emph{et~al.}
\newblock \bibinfo{title}{Pytorch: An imperative style, high-performance deep
  learning library}.
\newblock In \emph{\bibinfo{booktitle}{Advances in neural information
  processing systems}}, \bibinfo{pages}{8026--8037} (\bibinfo{year}{2019}).

\bibitem{fey2019fast}
\bibinfo{author}{Fey, M.} \& \bibinfo{author}{Lenssen, J.~E.}
\newblock \bibinfo{title}{Fast graph representation learning with pytorch
  geometric}.
\newblock \emph{\bibinfo{journal}{arXiv preprint arXiv:1903.02428}}
  (\bibinfo{year}{2019}).

\bibitem{Hutter_Iannuzzi_Schiffmann_VandeVondele_2014}
\bibinfo{author}{Hutter, J.}, \bibinfo{author}{Iannuzzi, M.},
  \bibinfo{author}{Schiffmann, F.} \& \bibinfo{author}{VandeVondele, J.}
\newblock \bibinfo{title}{cp2k: atomistic simulations of condensed matter
  systems}.
\newblock \emph{\bibinfo{journal}{WIREs Computational Molecular Science}}
  \textbf{\bibinfo{volume}{4}}, \bibinfo{pages}{15–25}
  (\bibinfo{year}{2014}).
\newblock \bibinfo{note}{00000}.

\bibitem{vasp1}
\bibinfo{author}{Kresse, G.} \& \bibinfo{author}{Hafner, J.}
\newblock \bibinfo{title}{Ab initiomolecular dynamics for liquid metals}.
\newblock \emph{\bibinfo{journal}{Physical Review B}}
  \textbf{\bibinfo{volume}{47}}, \bibinfo{pages}{558--561}
  (\bibinfo{year}{1993}).
\newblock \urlprefix\url{https://doi.org/10.1103/physrevb.47.558}.

\bibitem{vasp2}
\bibinfo{author}{Kresse, G.} \& \bibinfo{author}{Furthm\"{u}ller, J.}
\newblock \bibinfo{title}{Efficiency of ab-initio total energy calculations for
  metals and semiconductors using a plane-wave basis set}.
\newblock \emph{\bibinfo{journal}{Computational Materials Science}}
  \textbf{\bibinfo{volume}{6}}, \bibinfo{pages}{15--50} (\bibinfo{year}{1996}).
\newblock \urlprefix\url{https://doi.org/10.1016/0927-0256(96)00008-0}.

\bibitem{vasp3}
\bibinfo{author}{Kresse, G.} \& \bibinfo{author}{Furthm\"{u}ller, J.}
\newblock \bibinfo{title}{Efficient iterative schemes forab initiototal-energy
  calculations using a plane-wave basis set}.
\newblock \emph{\bibinfo{journal}{Physical Review B}}
  \textbf{\bibinfo{volume}{54}}, \bibinfo{pages}{11169--11186}
  (\bibinfo{year}{1996}).
\newblock \urlprefix\url{https://doi.org/10.1103/physrevb.54.11169}.

\bibitem{PBE}
\bibinfo{author}{Perdew, J.~P.}, \bibinfo{author}{Burke, K.} \&
  \bibinfo{author}{Ernzerhof, M.}
\newblock \bibinfo{title}{Generalized gradient approximation made simple}.
\newblock \emph{\bibinfo{journal}{Physical Review Letters}}
  \textbf{\bibinfo{volume}{77}}, \bibinfo{pages}{3865--3868}
  (\bibinfo{year}{1996}).
\newblock \urlprefix\url{https://doi.org/10.1103/physrevlett.77.3865}.

\bibitem{PAW}
\bibinfo{author}{Kresse, G.} \& \bibinfo{author}{Joubert, D.}
\newblock \bibinfo{title}{From ultrasoft pseudopotentials to the projector
  augmented-wave method}.
\newblock \emph{\bibinfo{journal}{Physical Review B}}
  \textbf{\bibinfo{volume}{59}}, \bibinfo{pages}{1758--1775}
  (\bibinfo{year}{1999}).
\newblock \urlprefix\url{https://doi.org/10.1103/physrevb.59.1758}.

\bibitem{kingma2014adam}
\bibinfo{author}{Kingma, D.~P.} \& \bibinfo{author}{Ba, J.}
\newblock \bibinfo{title}{Adam: A method for stochastic optimization}.
\newblock \emph{\bibinfo{journal}{arXiv preprint arXiv:1412.6980}}
  (\bibinfo{year}{2014}).

\bibitem{loshchilov2017decoupled}
\bibinfo{author}{Loshchilov, I.} \& \bibinfo{author}{Hutter, F.}
\newblock \bibinfo{title}{Decoupled weight decay regularization}.
\newblock \emph{\bibinfo{journal}{arXiv preprint arXiv:1711.05101}}
  (\bibinfo{year}{2017}).

\bibitem{reddi2019convergence}
\bibinfo{author}{Reddi, S.~J.}, \bibinfo{author}{Kale, S.} \&
  \bibinfo{author}{Kumar, S.}
\newblock \bibinfo{title}{On the convergence of adam and beyond}.
\newblock \emph{\bibinfo{journal}{arXiv preprint arXiv:1904.09237}}
  (\bibinfo{year}{2019}).

\end{thebibliography}

\section{Acknowledgements}
We thank Jonathan Vandermause, Cheol Woo Park, David Clark, Kostiantyn Lapchevskyi, Joshua Rackers, and Benjamin Kurt Miller for helpful discussions.\\

Work at Harvard was supported by Bosch Research and the Integrated Mesoscale Architectures for Sustainable Catalysis (IMASC), an Energy Frontier Research Center funded by the US Department of Energy (DOE), Office of Science, Office of Basic Energy Sciences under Award No. DE-SC0012573 and by Award No. DE-SC0022199. N.M. and B.K. are supported by a Multidisciplinary	University Research Initiative,	 sponsored	 by	 the Office	 of	 Naval	 Research, under Grant	 N00014-20-1-2418.\\

Work at Bosch Research was partially supported by ARPA-E Award No. DE-AR0000775 and used resources of the Oak Ridge Leadership Computing Facility at Oak Ridge National Laboratory, which is supported by the Office of Science of the Department of Energy under Contract DE-AC05-00OR22725.\\

T.E.S. was supported by the Laboratory Directed Research and Development Program of Lawrence Berkeley National Laboratory and the Center for Advanced Mathematics for Energy Research Applications, both under U.S. Department of Energy Contract No. DE-AC02-05CH11231.\\

A.M is supported by U.S. Department of Energy, Office of Science, Office of Advanced Scientific Computing Research, Computational Science Graduate Fellowship under Award Number(s) DE-SC0021110. \\

The authors acknowledge computing resources provided by the Harvard University FAS Division of Science Research Computing Group and by the Texas Advanced Computing Center (TACC) at The University of Texas at Austin under allocations DMR20009 and DMR20013.

\section{Author contributions}

S.B. initiated the project, conceived the NequIP model, implemented the software and conducted all software experiments under the guidance of B.K. A.M. contributed to the development of the model and the software implementation. L.S. created the data set and helped with MD simulations of formate/Cu, and contributed to the development of the model and its software implementation. M.G. contributed to the development of the model and the software implementation. J.P.M. contributed to analyzing the LiPS conductor results and implemented the thermostat for MD simulations together with S.B. M.K. generated the AIMD data set of Li\textsubscript{4}P\textsubscript{2}O\textsubscript{7}, wrote software for the analysis of MD results and contributed to the benchmarking on this system. N.M. wrote software for the estimation of diffusion coefficients and contributed to the interpretation of results. T.E.S. contributed to the conception of the model, guidance of computational experiments and software implementation. B.K. supervised the project from conception to design of experiments, implementation, theory, as well as analysis of data. All authors contributed to writing the manuscript. 

\section{Competing interests}

The authors declare no competing interests.

\newpage

\section{Appendix A: Long Molecular Dynamics Simulation of Li\textsubscript{4}P\textsubscript{2}O\textsubscript{7}}

\begin{figure*}[h!]
  \includegraphics[width=\linewidth]{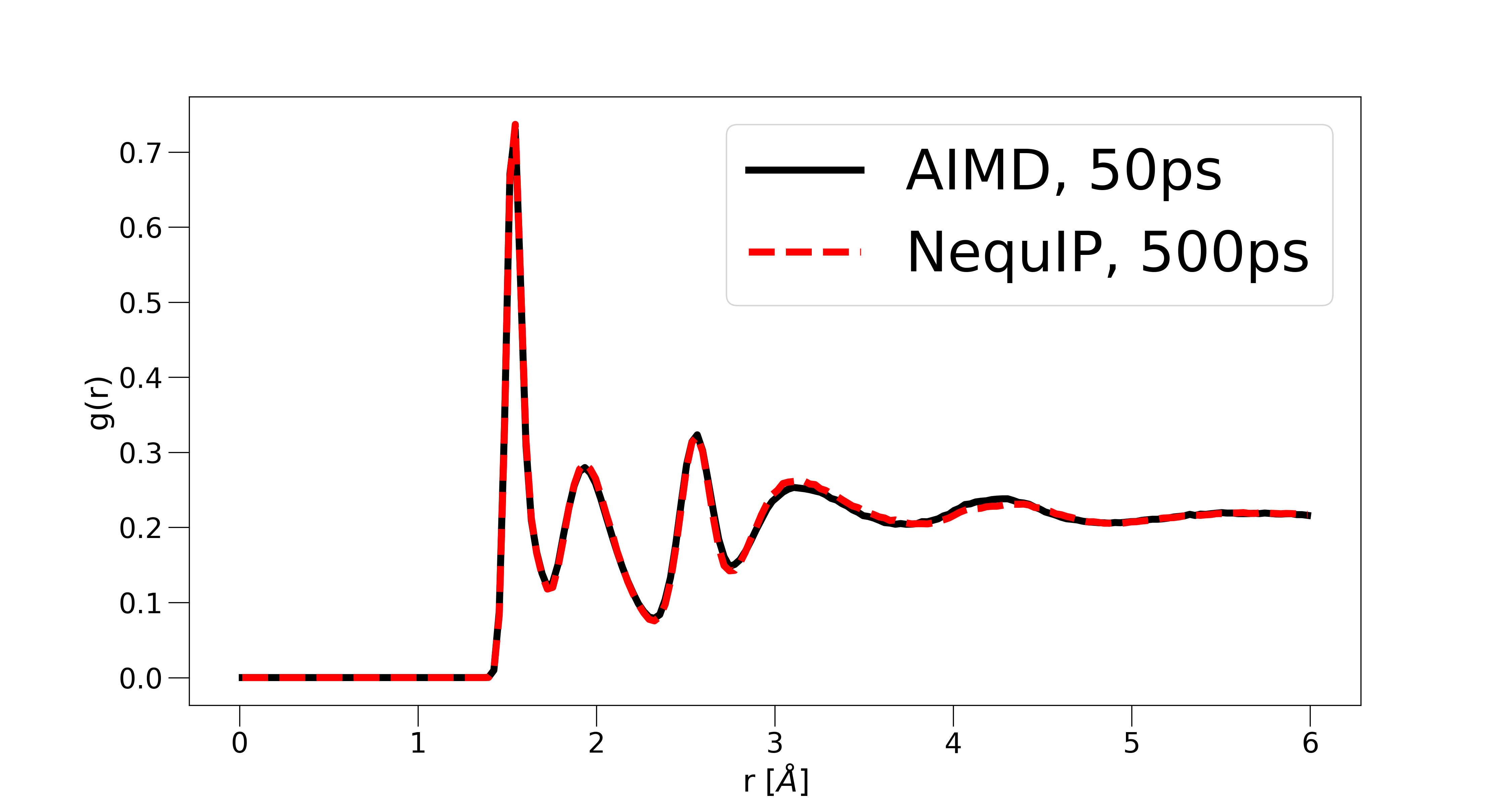}
  \caption{Comparison of the Radial Distribution of a simulation of Li\textsubscript{4}P\textsubscript{2}O\textsubscript{7} of length 500 ps driven by NequIP in comparison to a 50 ps simulation driven by AIMD.}
  \label{fig:long_lipo}
\end{figure*}

Figure \ref{fig:long_lipo} shows the Radial Distribution Function obtained from a MD simulation of simulation length 500ps compared to the AIMD simulation of 50ps. For both simulations, there first 10ps were not used in the computation of the RDF.

\section{Appendix B: Learning Curves}

\begin{figure*}[h!]
  \includegraphics[width=\linewidth]{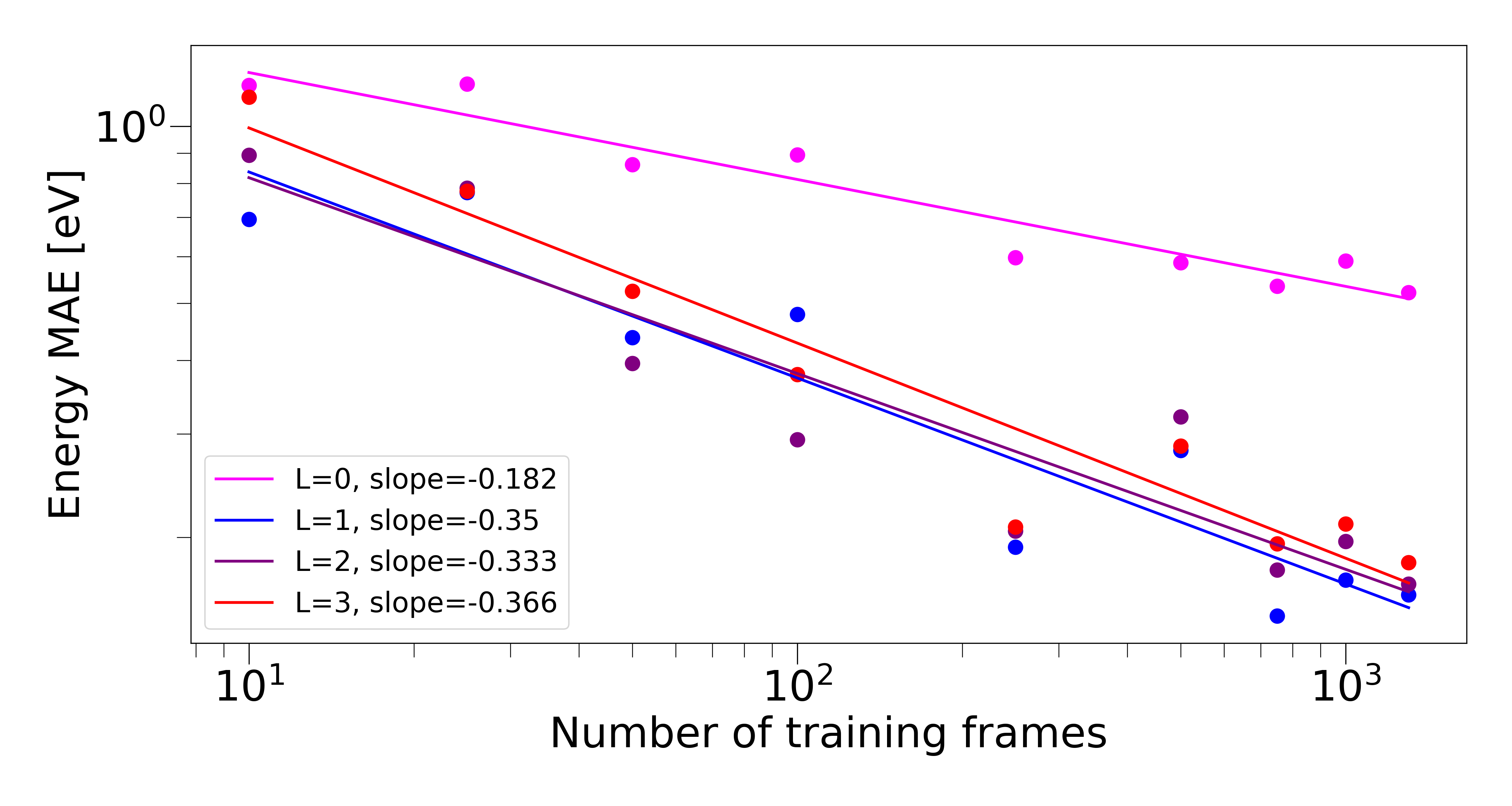}
  \caption{Log-log plot of the predictive error on the water data set from \cite{cheng2019ab} using NequIP with $l \in \{0, 1, 2, 3\}$ as a function of training set size, measured via the energy MAE.}
  \label{fig:water_e_data_eff}
\end{figure*}

\begin{figure*}[h!]
  \includegraphics[width=\linewidth]{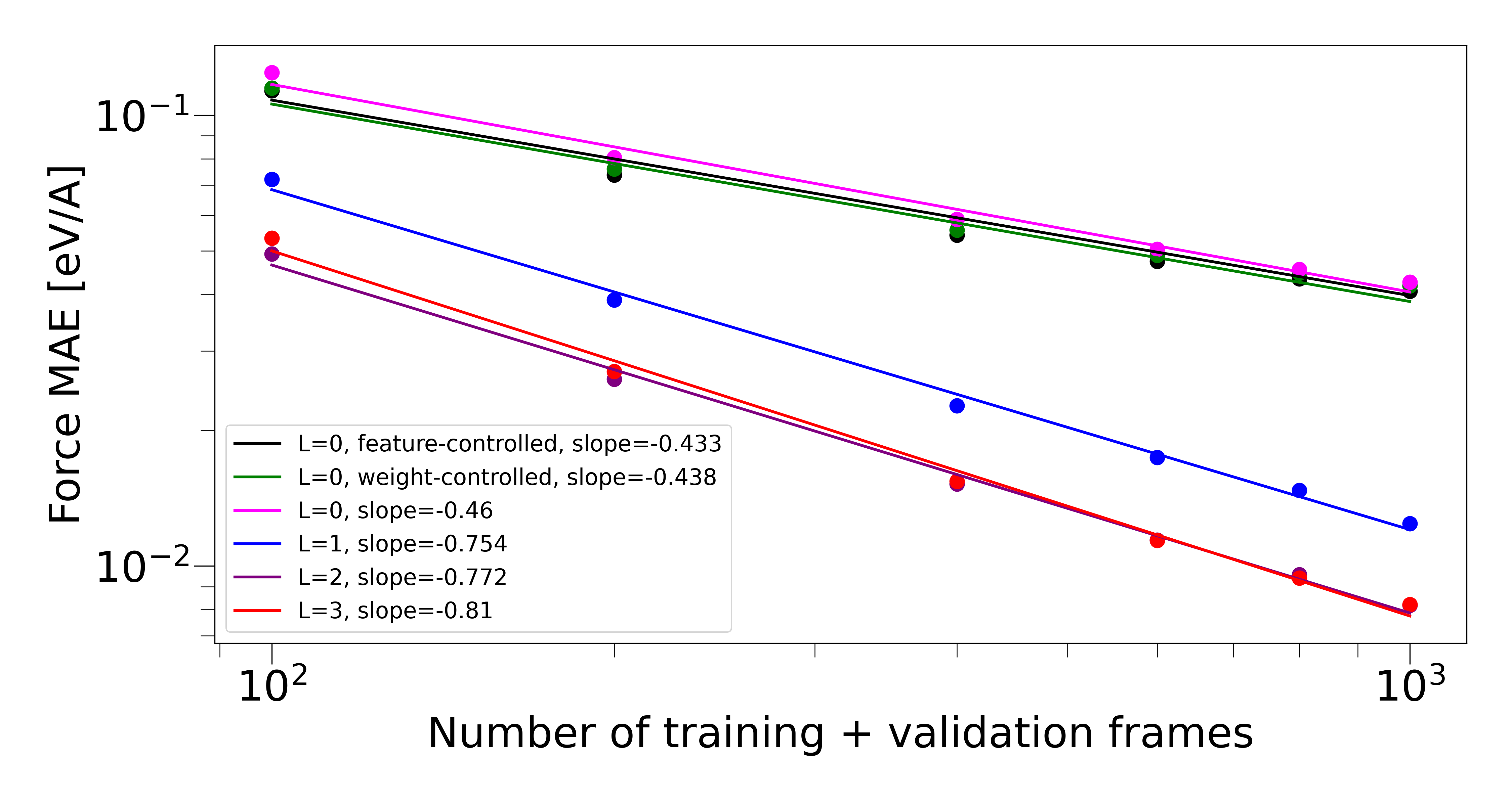}
  \caption{Log-log plot of the predictive error on the aspirin molecule in MD-17 using NequIP with $l \in \{0, 1, 2, 3\}$ as a function of data set size, measured via the force MAE. The plots also shows the weight- and feature-controlled version of NequIP.}
  \label{fig:water_f__data_eff_with_control}
\end{figure*}

Figure \ref{fig:water_e_data_eff} shows energy errors as a function of training set size on the data set from \cite{cheng2019ab}. Figure \ref{fig:water_f__data_eff_with_control} shows force errors as a function of training set size on the apsirin molecule in the MD-17 data set, together with weight- and feature-controlled version of the $l=0$ network. 

\section{Appendix C: Revised MD-17 data set }

\begin{figure*}[h!]
  \includegraphics[width=\linewidth]{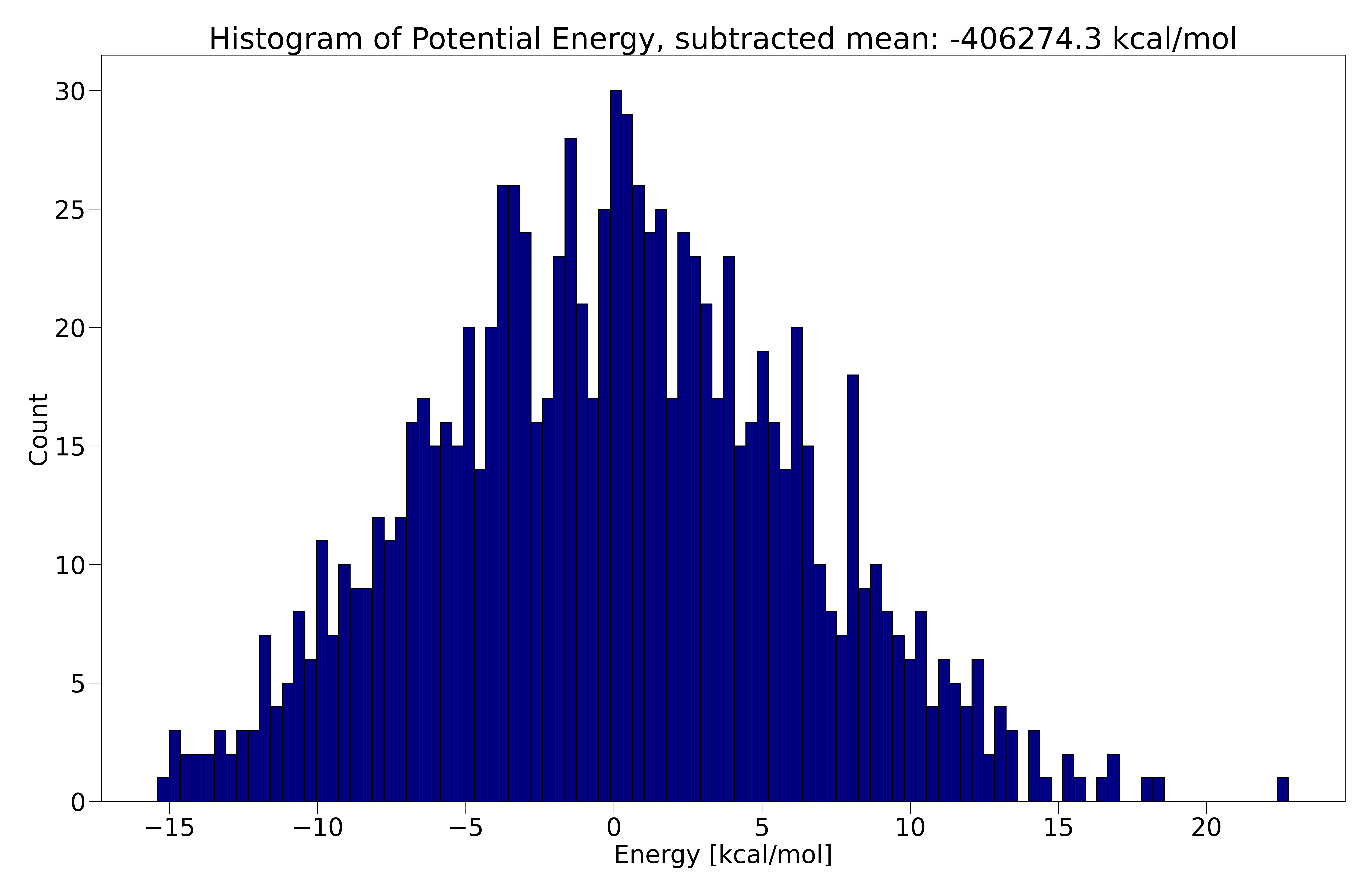}
  \caption{Histogram of potential energies of all structures used for training and validation for the aspirin molecule in the revised MD-17 data set. The mean energy was subtracted before plotting.}
  \label{fig:revmd17-pot-e-labels}
\end{figure*}

\begin{figure*}[h!]
  \includegraphics[width=\linewidth]{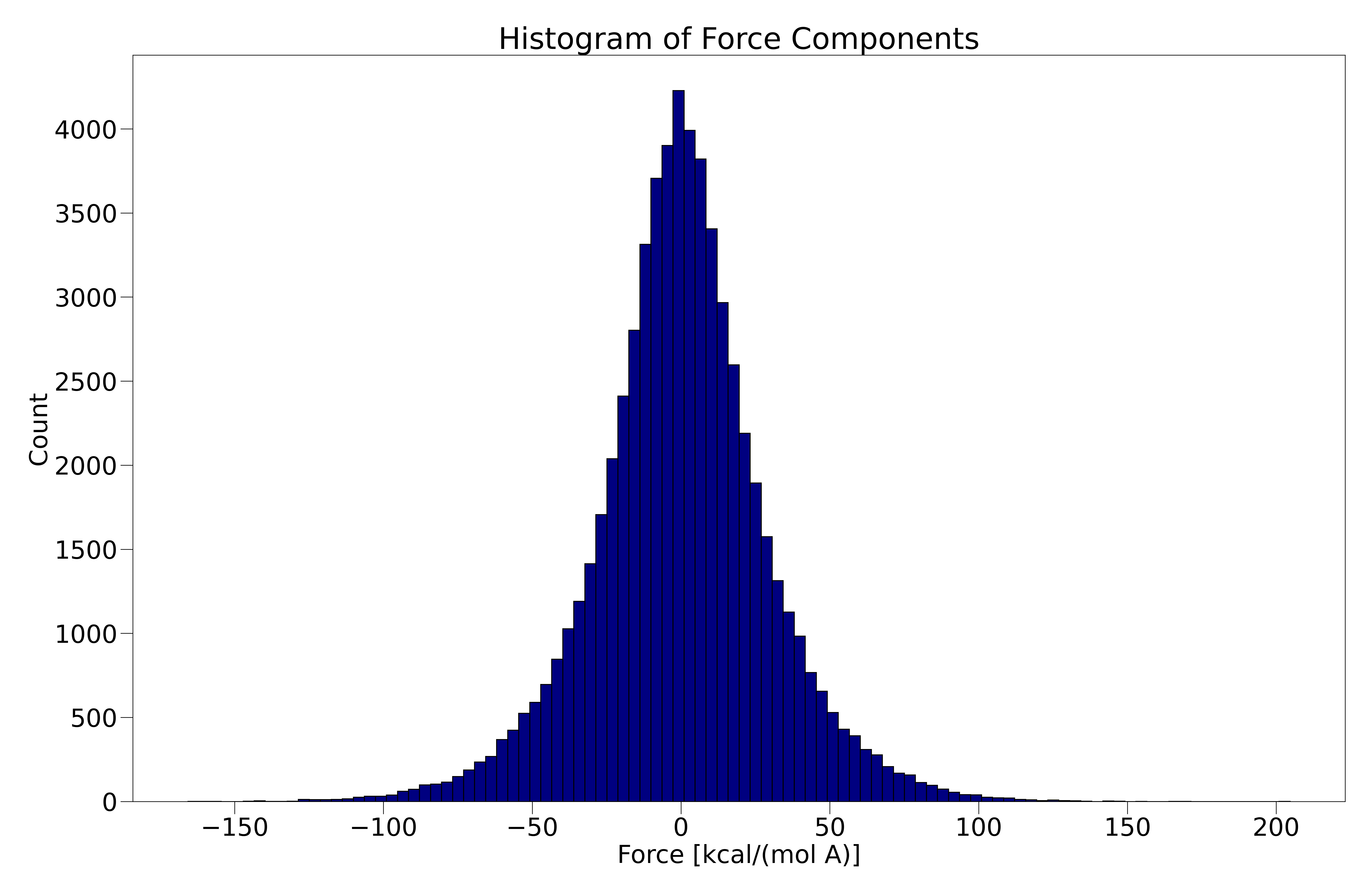}
  \caption{Histogram of force components of all structures used for training and validation for the aspirin molecule in the revised MD-17 data set.}
  \label{fig:revmd17-force-labels}
\end{figure*}

Figures \ref{fig:revmd17-pot-e-labels} and \ref{fig:revmd17-force-labels} show histograms of the energy and force labels on aspirin in the revised MD-17 data set, respectively.

\end{document}